\documentclass[a4paper,twocolumn,11pt,accepted=2025-08-22]{quantumarticle}

\pdfoutput=1

\usepackage[utf8]{inputenc}
\usepackage{amsfonts,amsmath,amssymb,graphicx,epstopdf,verbatim,dsfont,xcolor}
\usepackage[english]{babel}
\usepackage{subfigure}
\usepackage{graphicx}
\usepackage[colorlinks=true]{hyperref}
\usepackage{physics}
\usepackage{amsfonts}
\usepackage{mathtools}
\usepackage{booktabs} %
\usepackage{colortbl} %
\usepackage{xcolor}   %

\graphicspath{ {./figures/}}

 \definecolor{green_}{rgb}{0.2660, 0.6240, 0.1880}
 \definecolor{blue_}{rgb}{0, 0.4470, 0.7410}
 \definecolor{red_}{rgb}{0.7350, 0.0780, 0.1840}
 \definecolor{orange_}{rgb}{0.9290, 0.6940, 0.1250}
 \definecolor{purple_}{rgb}{0.4940, 0.1840, 0.5560}

\newcommand{\ColorX}{Red}
\newcommand{\colorX}{red}

\newcommand{\colorY}{green}

\newcommand{\colorZ}{blue}

\newcommand{\QuICS}{
Joint Center for Quantum Information and Computer Science, NIST/University of Maryland, College Park, Maryland 20742, USA}
\newcommand{\JQI}{
Joint Quantum Institute, NIST/University of Maryland, College Park, Maryland 20742, USA}

\begin{document}
\title{Fault-tolerant hyperbolic Floquet quantum error correcting codes}
\author{Ali~Fahimniya}
\email{fahim@umd.edu}
\affiliation{\QuICS}
\affiliation{\JQI}
\author{Hossein~Dehghani}
\affiliation{\QuICS}
\affiliation{\JQI}
\author{Kishor~Bharti}
\affiliation{\QuICS}
\affiliation{\JQI}
\affiliation{
Institute of High Performance Computing (IHPC), Agency for Science,
Technology and Research (A*STAR), 1 Fusionopolis Way,
\# 16-16 Connexis, Singapore 138632, Republic of Singapore}
\author{Sheryl~Mathew}
\affiliation{\QuICS}
\author{Alicia~J.~Koll\'ar}
\affiliation{\JQI}
\author{Alexey~V.~Gorshkov}
\affiliation{\QuICS}
\affiliation{\JQI}
\author{Michael~J.~Gullans}
\affiliation{\QuICS}

\begin{abstract}
   A central goal in quantum error correction is to reduce the overhead of fault-tolerant quantum computing by increasing noise thresholds and reducing the number of physical qubits required to sustain a logical qubit. We introduce a potential path towards this goal based on a family of dynamically generated quantum error correcting codes that we call ``hyperbolic Floquet codes.''  These codes are defined by a specific sequence of non-commuting two-body measurements arranged periodically in time that stabilize a topological code on a hyperbolic manifold with negative curvature. We focus on a family of lattices for $n$ qubits that, according to our prescription that defines the code, provably achieve a finite encoding rate $(1/8+2/n)$ while still requiring only two-body measurements. Similar to hyperbolic surface codes, the distance of the code at each time-step scales at most logarithmically in $n$. The family of lattices we choose indicates that this scaling is achievable in practice. We develop and benchmark an efficient matching-based decoder that provides evidence of a threshold near $0.1\, \%$ in a phenomenological noise model and $0.25\, \%$ in an entangling measurements noise model. Utilizing weight-two check operators and a qubit connectivity of 3, one of our hyperbolic Floquet codes uses 400 physical qubits to encode 52 logical qubits with a code distance of 8, i.e., it is a $[[400,52,8]]$ code. At small error rates, comparable logical error suppression to this code requires 5x as many physical qubits (1924) when using the honeycomb Floquet code with the same noise model and decoder.    
\end{abstract}

\maketitle

\section{Introduction}
Large-scale quantum computing requires quantum error correction and fault-tolerance protocols to control the growth of errors.  
The surface code~\cite{kitaev_quantum_1997,kitaev_faulttolerant_2003,dennis_topological_2002} is a promising route towards fault-tolerant quantum computing because of its high tolerance for noise and geometrically local implementation. These features have led to considerable interest in realizing small instances of the surface code, as demonstrated in recent experiments~\cite{takita_demonstration_2016,marques_logicalqubit_2022,krinner_realizing_2022,zhao_realization_2022,acharya_suppressing_2023,bluvstein_quantum_2022,cong_enhancing_2024,semeghini_probing_2021}. 
In the conventional surface code, both the parity-check weight and the qubit degree are relatively low, specifically, each is only four. This means that each qubit is involved in just four parity checks, which are operations that help identify and correct errors. However, a drawback of the surface code is that it requires a large number of physical qubits to represent each logical qubit~\cite{bravyi_highthreshold_2024}. 
To improve on the performance of the surface code in general fault-tolerant protocols, one can attempt to optimize the syndrome extraction circuit by using weight-two check operators~\cite{hastings_dynamically_2021,chao_optimization_2020,gidney_pair_2023}, tailor the error correction scheme to specific noise models \cite{tuckett_tailoring_2019,bonillaataides_xzzx_2021,dua_clifforddeformed_2024,wu_erasure_2022}, or reduce the number of physical qubits required to achieve a certain number of logical qubits with a given code distance \cite{gottesman_faulttolerant_2014}.

Current schemes either support weight-two check operators or maintain a constant ratio of logical to physical qubits, but not both. For instance, recent research on quantum Low-Density Parity-Check (LDPC) codes~\cite{bravyi_highthreshold_2024} suggests that these codes can encode $12$ logical qubits using $288$ physical qubits. This compares favorably against surface codes which would require $4000$ physical qubits to encode $12$ logical qubits for similar error suppression. In an LDPC-type quantum error-correcting code, each check operator acts on only a few qubits, and each qubit is part of only a few checks~\cite{gottesman_faulttolerant_2014,tremblay_constantoverhead_2022,breuckmann_quantum_2021,panteleev_asymptotically_2022,leverrier_quantum_2022,bravyi_highthreshold_2024,xu_constantoverhead_2024a}. However, these particular quantum LDPC codes~\cite{bravyi_highthreshold_2024} necessitate the use of weight-six checks, rendering their implementation challenging with current devices. Hyperbolic quantum codes~\cite{luo_z2systolic_2002,zemor_cayley_2009,breuckmann_constructions_2016, breuckmann_hyperbolic_2017} that employ tessellations of closed hyperbolic surfaces have been proposed as a solution for achieving finite encoding rate with weight-three check operators. Recently, there has been considerable attention directed toward a new class of codes known as Floquet codes~\cite{hastings_dynamically_2021, gidney_faulttolerant_2021,vuillot_planar_2021,ellison_floquet_2023,davydova_floquet_2023,kesselring_anyon_2024,davydova_quantum_2024,aasen_adiabatic_2022,sullivan_floquet_2023}. These codes feature weight-two check operators but come with the limitation of a vanishing encoding rate; thus, raising  the question of whether it is possible to achieve a finite encoding rate using weight-two check operators instead. 

In this work, we introduce a new family of codes, termed as hyperbolic Floquet codes, with a finite encoding rate of $1/8 + 2/n$  for $n$ physical qubits and weight-two check operators. For example, with $400$ physical qubits, our scheme can encode $52$ logical qubits.
See Table~\ref{table:code_family} for a summary of some of our hyperbolic Floquet quantum error correcting codes. The hyperbolic Floquet codes inherit some of the best features from quantum hyperbolic codes~\cite{luo_z2systolic_2002,zemor_cayley_2009,breuckmann_constructions_2016, breuckmann_hyperbolic_2017} and Floquet codes~\cite{hastings_dynamically_2021}, although they give up on geometric locality. Similar to Floquet codes, our check operators have a weight of two, while the encoding rate is finite, akin to that of quantum hyperbolic codes. The implementation of our code demands a reduced qubit connectivity of $3$, in contrast to the recently introduced quantum LDPC codes~\cite{bravyi_highthreshold_2024} which may require a connectivity of $6$. 

Our approach excels in terms of the weight of the check operators, qubit connectivity, and encoding rate. However, a trade-off is that code distance ($d$) can scale at most logarithmically in the number of physical qubits. We find a family of hyperbolic lattices of increasing size whose associated hyperbolic Floquet codes achieve a distance scaling consistent with this expectation. 
We develop an efficient decoder based on matching of syndrome pairs similar to the surface code. Employing two different noise models%
, we see evidence for %
thresholds around $0.1\, \%$ and $0.25\, \%$. The finite encoding rate implies that a total logical error suppression can be reached with an asymptotically vanishing fraction of the physical qubits that would be required using zero-rate topological codes. To further substantiate this scaling argument, we provide a detailed comparison of the logical failure rate of our codes against many independent copies of the honeycomb Floquet code under the same noise models and decoder.

The architecture of hyperbolic Floquet codes is elaborated upon in Section~\ref{sec:hyperbolic_Floquet_code}, where we explore various hyperbolic lattices, measurement protocols and logical operators. The properties related to error detection and correction are described %
in Section~\ref{sec:decoding}. In Section~\ref{sec:results}, we present numerical results of logical error rate suppression and evidence for a threshold, as well as compare hyperbolic Floquet codes in performance to honeycomb Floquet codes within the same noise model and decoder. Finally, we conclude with discussion and several open problems in Section~\ref{sec:discussion}.

\section{The Hyperbolic Floquet Code} \label{sec:hyperbolic_Floquet_code}
\begin{figure}[bt]
	\centering
	\includegraphics[width=\linewidth]{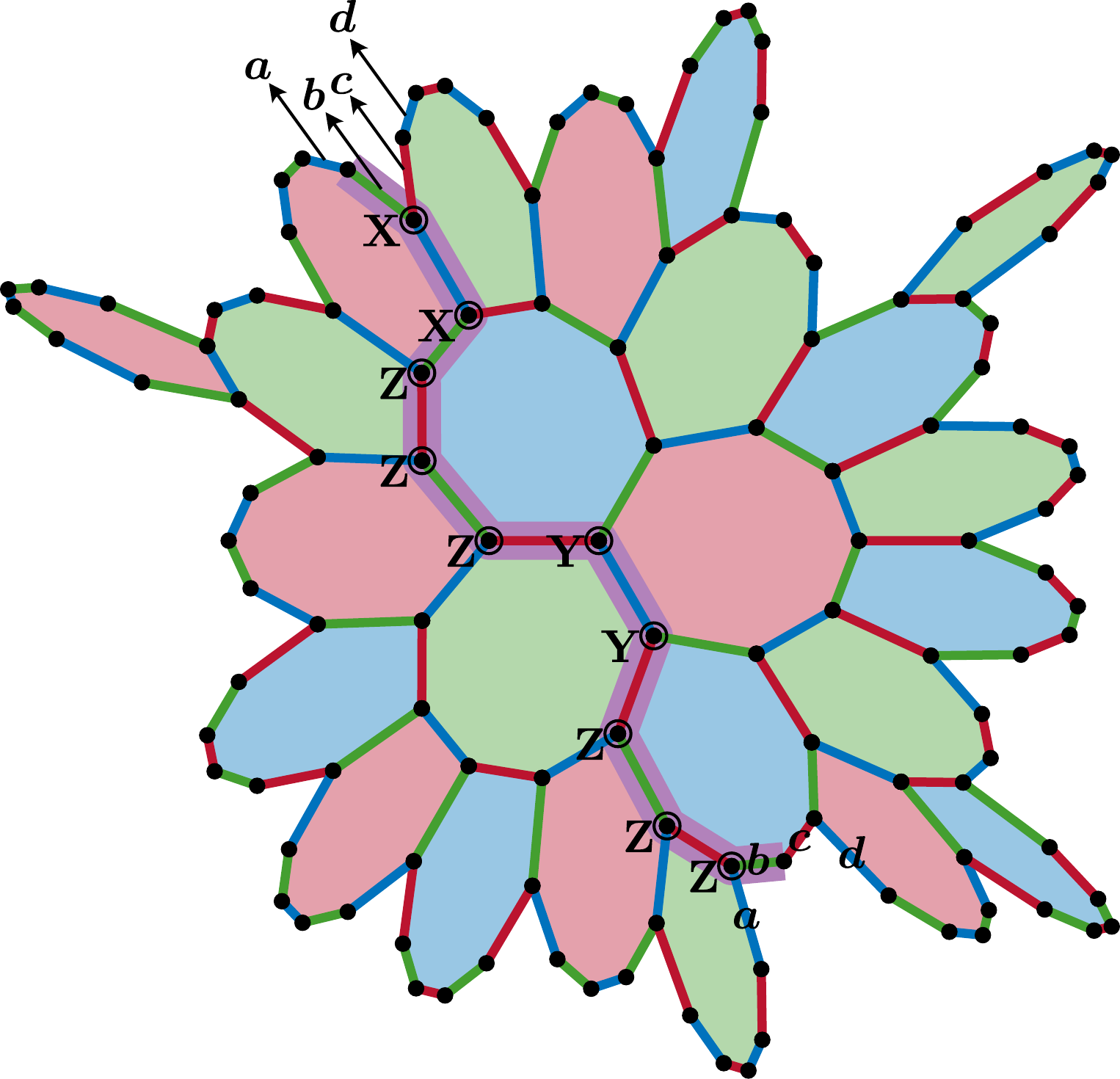}
	\caption{A hyperbolic Floquet code with 64 physical qubits.
	The code is based on an octagonal (\{8, 3\}) lattice, which is displayed using polar coordinates of its vertices. The 64 physical qubits are located at the vertices of the lattice. Identification of the edges at the boundary makes the lattice a periodic tessellation of a genus-5 manifold. Four such periodic boundary condition identifications are marked by letters $a$-$d$. The octagonal faces are three-colored such that the adjacent faces have different colors. Each edge inherits the color from the two faces it would enter if elongated. \ColorX, \colorY, and \colorZ{} edges correspond to XX, YY, and ZZ measurement checks, respectively. The \colorY{} edges at the ends of the highlighted purple path are identified as a periodic boundary condition, marked with letter $b$. The purple path therefore is a closed path consisting of 10 unique vertices, marked by circles. This closed path hosts a weight-ten noncontractible loop operator, Pauli operators of which are denoted next to its vertices.}
	\label{fig:lattice}
\end{figure}

\paragraph*{Lattice and Measurement Protocol.}
Hyperbolic Floquet codes utilize regular tessellations of negatively curved planes, employing even-sided polygons as building blocks with a degree of three for all vertices. The physical qubits are located at the vertices of the lattice. This code relies on a three-coloring of the polygons, and thus, it is required that they are even-sided. Figure~\ref{fig:lattice} depicts a face three-colored lattice of octagons that is the basis for a 64-qubit hyperbolic Floquet code. This lattice is of the form \{8, 3\} using Sch\"afli symbols~\cite{a.blatov_vertex_2010}, meaning that 3 octagons (8-gons) meet at each vertex. We refer to this lattice as the \textit{octagonal lattice}, and employ it in the development of hyperbolic Floquet codes. Hyperbolic Floquet codes can also be derived from lattices with decagons or higher even-sided polygons, though such codes fall outside of the scope of this paper's in-depth analysis.

While the octagonal lattice is regular and all vertices are equivalent, the projection of this lattice from the negatively-curved hyperbolic plane onto a flat plane, as in Fig.~\ref{fig:lattice}, results in distortions in the shapes of the displayed octagons. We employ the polar coordinates of the vertices for this projection. That is, each vertex's projected coordinates are obtained based on its distance from a reference vertex using the hyperbolic metric and its orientation with respect to a reference direction. The depicted lattice is a closed lattice, with complex periodic boundary conditions, tessellating a genus-5 manifold. The periodic boundary conditions identify each edge that does not seem to have two adjacent octagons with another such edge. Four of these boundary condition identifications are labeled with the letters $a$, $b$, $c$, and $d$. Edges with the same letter are identified as equivalent, thus representing the same edge. Similarly, the four vertices at the ends of the paired edges with matching labels are identified and represent only two distinct vertices.

\begin{figure}[t!]
	\centering
	\includegraphics[width=\linewidth]{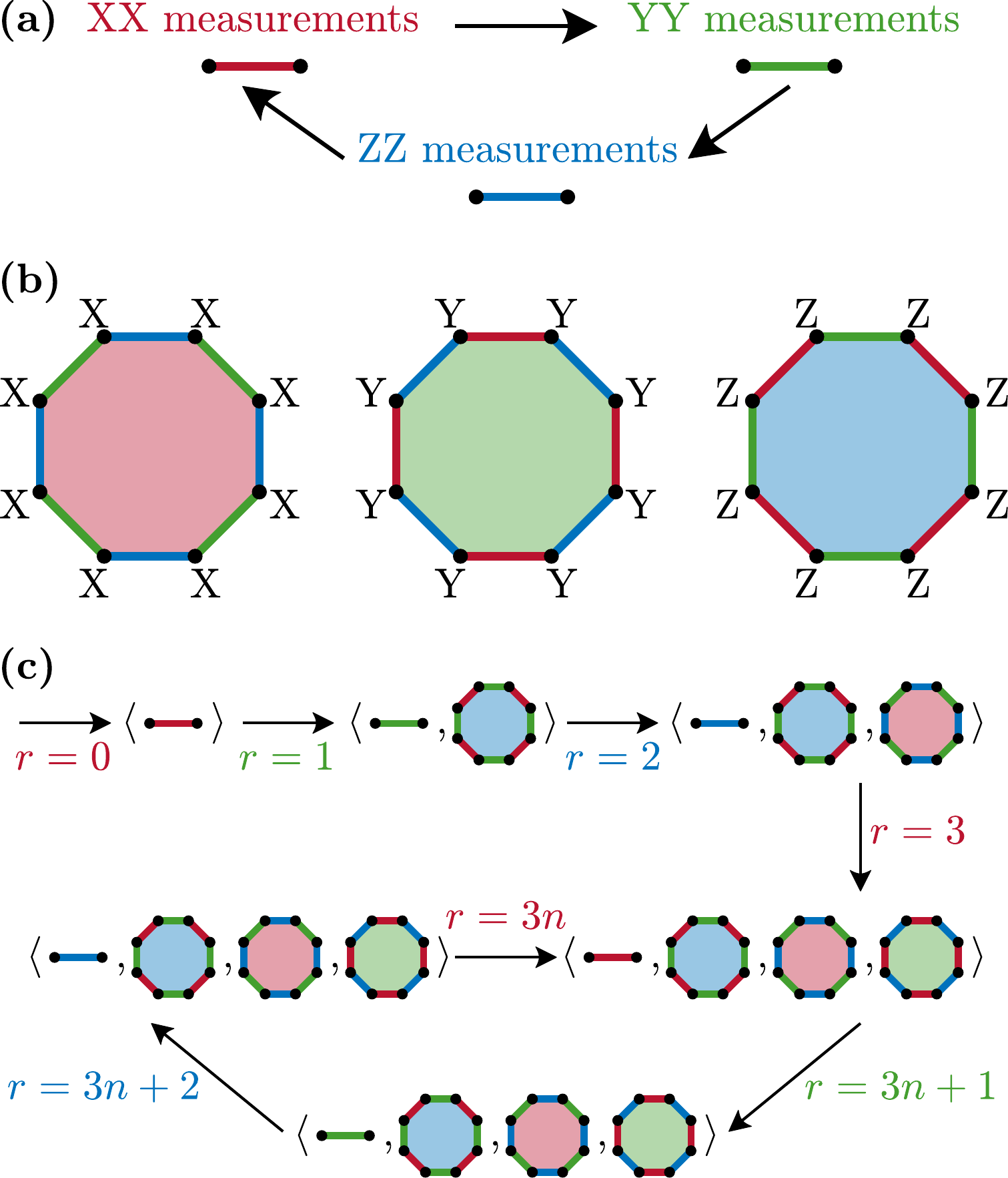}
	\caption{Measurement protocol and plaquette operators of hyperbolic Floquet codes.
    (a) One cycle of the Floquet code's measurement protocol. Each cycle consists of three measurement rounds, each round carried out on XX, YY, and ZZ checks, corresponding to \colorX, \colorY, and \colorZ{} edges in Fig.~\ref{fig:lattice}, respectively. In each measurement, the weight-two Pauli operator of the two qubits at the ends of that edge are measured.
	(b) Plaquette operators that are products of checks along the edges of each octagon, for \colorX, \colorY, and \colorZ{} plaquettes. They commute with all check operators on the lattice, and their value can be inferred using the measurement outcomes of the check operators.
    (c) The instantaneous stabilizer group (ISG) after each measurement round expressed as generated by its generators after each measurement round. The generators can be check operators (panel a) or plaquette operators (panel b). Each check or plaquette operator represents all similar operators with the same color up to a sign.}
	\label{fig:protocol}
\end{figure}

The edge colors of the octagonal lattice in Fig.~\ref{fig:lattice} are derived from the face colors, such that each edge inherits its color from the two octagonal faces that it connects and would extend into if elongated. In the Floquet code, these three-colored edges constitute the repetitive rounds of two-body check measurements of the physical qubits of the edges. At rounds ${r = 3n}$ (${r = 3n+1}$, ${r = 3n+2}$) with nonnegative integers $n$, all \colorX{} (\colorY, \colorZ) checks are measured, as illustrated in Fig.~\ref{fig:protocol}(a). Specifically, \colorX, \colorY, and \colorZ{} checks correspond to measurements of XX, YY, and ZZ Pauli operators on the pairs of physical qubits at the ends of the respective edges. Although any two such measurement checks do not necessarily commute, all the checks within a single round do commute and can be considered to be measured simultaneously.

It is noteworthy that the original honeycomb Floquet code described by Hastings and Haah~\cite{hastings_dynamically_2021} requires a second edge labeling based on edge orientations, which dictates the type of two-body measurement. However, as later indicated by Gidney et al.~\cite{gidney_faulttolerant_2021}, this second labeling is redundant, as the edge colors alone can determine the measurement type, yielding the same code. These two code definitions can be mapped onto one another through local Bloch rotations of the physical qubits. In this work, we adopt the version requiring solely the edge three-coloring to define hyperbolic Floquet codes. This approach is advantageous because edge orientation is inherently ambiguous in hyperbolic space.

\vspace{0.1in}
\paragraph*{Instantaneous Stabilizer Group.}
In this section, we assume errorless evolution of the quantum state of the physical qubits. We further consider that this state starts as a maximally mixed state prior to the round $r=0$ of measurements. Then, after each measurement round, we examine the group of all Pauli operators that have an eigenvalue $+1$ on the current state of the physical qubits, i.e.,~stabilize the current state. This group is termed the \textit{instantaneous stabilizer group} (ISG), and before any measurements, consists of solely the identity operator. The check measurements, however, change the structure of the ISG. Figure~\ref{fig:protocol}(c) shows the evolution of the ISG through measurement rounds.

After rounds ${r = 3n}$ (${r = 3n+1}$, ${r = 3n+2}$), each XX check (YY check, ZZ check) becomes a generator of the ISG, up to a sign determined by the respective measurement outcome (i.e.,~if the measurement result of a check is ${-1}$, the negative of that check joins the ISG). For every one of these two-body generators of the ISG, two checks in the subsequent measurement round anticommute with it. As a result, these two-body instantaneous stabilizer generators no longer stabilize the state following the next round. For example, consider the four \colorX{} (i.e.,\ XX) edges around the \colorZ{} face in Fig.~\ref{fig:protocol}(b). After round ${r=3n}$, the two-body checks of these edges, up to their measurement results' signs, are members of the ISG. The upcoming measurement round, ${r=3n+1}$, is \colorY{} (i.e.,\ YY) checks. Each \colorX{} check anticommutes with two \colorY{} checks from of the \colorZ{} face, and hence, will be removed from the ISG upon measurement of any of those \colorY{} checks. After round ${r=3n+1}$, each of these \colorY{} checks, up to its measurement outcome sign, will be in the ISG.

Although the two-body checks are removed from the ISG following the next measurement round, certain products of them commute with the upcoming measurement checks, and therefore, stay in the ISG. In the example of the \colorZ{} face and \colorX{} checks in Fig.~\ref{fig:protocol}(b), the product of all four \colorX{} checks commutes with any \colorY{} check, ensuring its membership in the ISG after round ${r=3n+1}$. The product of these four \colorX{} checks can be multiplied with the product of the four \colorY{} checks of that \colorZ{} face, deriving a weight-eight plaquette operator. This operator, after round ${r=3n+1}$, is part of the ISG. Since it is equal to a product of X and Y Paulis for each qubit, it is equivalent to a product of Z Paulis (up to a phase). In a similar vein, plaquette operators for \colorX{} and \colorY{} faces are products of X and Y Paulis, respectively. The plaquette operators are depicted in Fig.~\ref{fig:protocol}(b).

Each plaquette operator [Fig.~\ref{fig:protocol}(b)] shares the same Pauli operator as the two-body checks intersecting it at a single vertex. Conversely, it has a different Pauli operator than the two-body checks intersecting it at two vertices. Consequently, these plaquette operators commute with all check operators. When measured (during rounds $r=2$, $r=3$, and $r=4$ corresponding to \colorZ, \colorX, and \colorY{} faces), they will join the ISG, up to a sign derived from the measurements. Beyond round $r=4$, the ISG mod phases reaches a periodic steady state with the logical qubits in a mixed state (i.e., stable against the non-commuting measurements), consistently generated by all plaquette operators and the most recent measurement checks, each acquiring a sign from the measurement outcomes.

\begin{figure*}[bt]
	\centering
	\includegraphics[width=\linewidth]{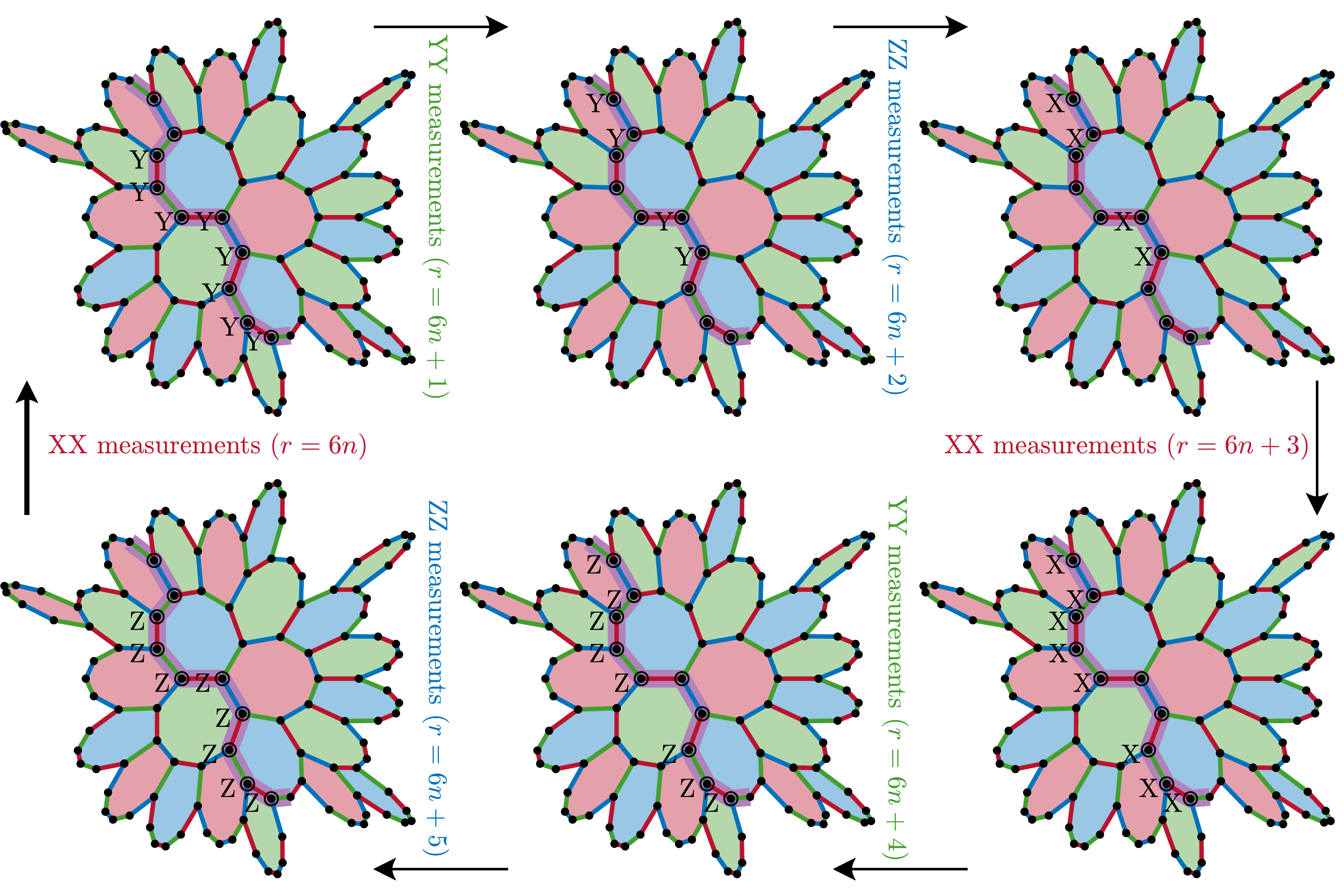}
	\caption{Evolution of a logical operator in the 64-qubit hyperbolic Floquet code through measurement rounds. The logical operator is contained within the noncontractible loop operator (see Fig.~\ref{fig:lattice}) of a closed path (highlighted in purple). The nonidentity Pauli operators of the logical operator are denoted next to the physical qubits. The logical operator commutes with the plaquette operators, the most recent measurement checks, and the upcoming measurement checks. After each round of measurements, the logical operator's expression is updated by being multiplied with the measured checks along the purple path. The update procedure results in changing a type one logical operator to a type two, and vice versa. Consequently, a periodicity of six for the logical operator's expression emerges. The weight of the logical operator varies through rounds, with a minimum value of four.}
	\label{fig:logical_evolution}
\end{figure*}

\vspace{0.1in}
\paragraph*{Logical Operators.}
Logical operators are the operators that commute with the ISG but are not part of it. Commuting with any two consecutive instances of the ISG is required for the logical information to not be destroyed by the measurement rounds~\cite{aasen_measurement_2023}. However, operators that commute with the ISG but are always a part of it, i.e., the plaquette operators and their products, have determined values and cannot be used for restoring information about the logical qubits. The logical operators of the hyperbolic Floquet code are contained within noncontractible closed paths of the hyperbolic lattice. These are the paths that, utilizing the periodic boundary conditions, wrap around the lattice. One closed path of the hyperbolic lattice in Fig.~\ref{fig:lattice} is highlighted in purple.

Just as the product of two-body checks around an octagon commutes with all two-body checks, the product along any loop does the same. We will call these \textit{loop operators}. The contractible loop operators can be expressed as the product of the plaquette operators within them, and they are part of the ISG beyond round $r=4$. On the other hand, the noncontractible loop operators will never integrate into the ISG. The closed path of one such loop is highlighted with purple in Fig.~\ref{fig:lattice}. The displayed periodic boundary condition identifications (edges labeled with letters $a$ to $d$) serve as a reference to confirm that the purple path is indeed closed. The lattice's periodic boundary conditions identify the two \colorY{} edges at the ends of the portrayed path, making the two vertices at each end non-unique. The distinct vertices of the weight-ten loop operator are marked with a circle, and the adjacent Pauli operators represent the components of the loop operator. It is noteworthy that the Pauli operator of each physical qubit is determined by its outgoing edge that is not part of the loop; specifically, X, Y, or Z Paulis for \colorX, \colorY, or \colorZ{} outgoing edges, respectively. Noncontractible loop operators comprise the Floquet code's logical operators. As we will discuss below, the number of noncontractible loop operators is twice the genus of the manifold.

Each noncontractible loop operator can be decomposed into the product of two operators, where both individually commute with the elements of the ISG. This decomposition, akin to the ISG, depends on the most recent measurement round. Assume that measurements of the \colorX{} check operators (XX measurements) have just been carried out (round $r=3n$). The next measurement round is the \colorY{} checks (YY measurements). The two operators that commute with the ISG elements and make a noncontractible loop operator are the product of X Paulis on the vertices of the \colorY{} edges along its closed path (bottom right panel in Fig.~\ref{fig:logical_evolution}) and the product of Y Paulis on the vertices of the \colorX{} edges along it (top left panel in Fig.~\ref{fig:logical_evolution}). These two constructions are generalizable to any measurement round as \textit{the most recent measurement Paulis on the edges of the upcoming round} and \textit{the upcoming measurement Paulis on the edges of the most recent round}, and hereinafter will be referred to as the \textit{type one} and \textit{type two} logical operators. %

Given that, after XX measurements, the type one logical operator has only X Paulis and the type two has two Paulis on each XX check, they commute with the two-body XX checks in the ISG. They also commute with plaquette operators within the ISG. Taking the type one operator as an example, each \colorX{} and \colorZ{} plaquette shares an integer number of \colorY{} edges with the closed path. This means it has an even number of non-identity Paulis in common with the first operator. Moreover, since the path is closed and always leaves any octagon that it enters, an even number of its \colorY{} edges are incident at any given \colorY{} octagon. Therefore, there are again an even number of mutual non-identity Paulis between the operator and any \colorY{} plaquette operator. Similar analysis can be carried out for the type two operator, establishing that both operators commute with the ISG. Further, these operators are independent operators and one of them cannot be obtained by multiplying the other one with some elements of the ISG. These are the two logical operators defined on the noncontractible loop.

So far, the constructed logical operators have been found to commute with the ISG after XX measurements. Notably, this construction ensures that they also commute with the upcoming measurements, specifically YY checks on \colorY{} edges. To elaborate, the type one logical operator features two X Paulis on any \colorY{} edge along its path and the type two operator is made of Y Paulis. Therefore, YY checks can be measured without destroying these logical operators. However, after YY measurements, some of the upcoming ZZ checks anticommute with the current construction of these logical operators and their measurement is going to affect the logical operators. Nonetheless, before ZZ measurements, we can update the expression of the logical operators by multiplying them with the measured YY checks along their noncontractible loops. The measurement outcomes of these YY checks must be factored into the sign of the updated logical operators.

Consider the type two logical operator (top left panel of Fig.~\ref{fig:logical_evolution}). When we multiply this operator by YY checks on the closed path, the Y Paulis shift to the \colorZ{} edges (top middle panel of Fig.~\ref{fig:logical_evolution}). Note that this is a type one logical operator after YY (\colorY) and prior to ZZ (\colorZ) measurements. Following the next measurement round (ZZ measurements), we multiply this operator with the measured ZZ checks along the path, making it comprised of X Paulis on the \colorZ edges (top right panel of Fig.~\ref{fig:logical_evolution}). This is in turn a type two logical operator after ZZ and prior to XX measurements. As seen in Fig.~\ref{fig:logical_evolution}, the logical operator switches its type with each measurement round, and reverts to its original expression after six rounds.

In Ref.~\cite{hastings_dynamically_2021}, the noncontractible loop operators are called the inner logical operators, while the type one and type two logical operators are equivalent to the electric and magnetic outer logical operators defined therein. As the noncontractible loop operators are equal to the product of two logical operators, their existence does not add to the number of independent logical operators, and hence, the number of logical qubits of a hyperbolic Floquet code. In order to emphasize this, we do not use the term logical operators in referring to them.

\vspace{0.1in}
\paragraph*{Code Family.}
Now we focus on finding the lattices that give rise to the particular codes and calculate their code parameters. Each code is based on a periodic octagonal lattice. To identify lattices, we turn to the database of symmetric connected graphs of degree three up to 10,000 vertices, provided in~\cite{conder2002trivalent}\footnote{Reference~\cite{conder2002trivalent} includes methodology and graphs up to 768 vertices. Graphs with up to 10,000 vertices can be found \href{https://www.math.auckland.ac.nz/~conder/symmcubic10000list.txt}{here}.}. Within this database, the graphs with a \textit{girth} of eight qualify as periodic octagonal lattices. These lattices are symmetric (arc-transitive) graphs: for any pair of edges, there exists an automorphism of the graph that maps one edge to the other.

The number of vertices of each lattice determines the number of the physical qubits, $n$, for the corresponding code. The number of logical qubits, $k$, on the other hand, is determined by the genus of the manifold that the periodic lattice tessellates. In order to find the relation between the genus of this manifold and $n$, we need to look at the crystallography structure of the octagonal lattice~\cite{boettcher_crystallography_2022}.

An octagonal lattice has a Bravais unit cell comprising 16 vertices, which tessellates the surface of a genus-2 manifold (see Fig.~12 in Ref.~\cite{boettcher_crystallography_2022}). Every additional set of 16 vertices in a given octagonal lattice corresponds to one unit cell and cover an area equal to $-4\pi$ (in units of negative curvature) on the hyperbolic plane. Thus, the genus of the manifold covered by the entire lattice is given by:
\begin{equation}\label{eq:genus}
	g = \frac{n}{16} + 1.
\end{equation}

\begin{table}
    \centering
    \resizebox{\linewidth}{!}{
    \begin{tabular}{cccc}
        \toprule
        \textbf{Code} & \textbf{Unit Cells \#} & \textbf{Genus} & \textbf{[[\textit n, \textit k, \textit d]]} \\
        \midrule        
        \textcolor{red_}{\textbf{H16}} & 1 & 2 & [[16, 4, 2]] \\
        \textcolor{green_}{\textbf{H64}} & 4 & 5 & [[64, 10, 4]] \\
        \textcolor{blue_}{\textbf{H144}} & 9 & 10 & [[144, 20, 6]] \\
        \textcolor{orange_}{\textbf{H400}} & 25 & 26 & [[400, 52, 8]] \\
        \textcolor{purple_}{\textbf{H2160}} & 135 & 136 & [[2160, 272, 10]] \\
        \bottomrule
    \end{tabular}}
    \caption{Family of hyperbolic octagonal Floquet codes with the fewest number of physical qubits for each code distance. Each code is denoted by a prefix `H' followed by the number of its physical qubits. The layout of the physical qubits corresponds to a tessellation of a negatively curved manifold with one Bravais unit cell per each 16 physical qubits. The number of handles (genus) of this manifold is one more than the number of unit cells [Eq.~\eqref{eq:genus}]. The number of the logical qubits, $k$, is twice the genus [Eq.~\eqref{eq:logicals}], and the code distance, $d$, is determined by finding the logical operator with the lowest minimum weight.}
    \label{table:code_family}
\end{table}

Having a genus that grows with the number of physical qubits, hyperbolic error correcting codes exhibit a finite encoding rate. This stems from the fact that a manifold's number of independent noncontractible closed paths equals twice its genus. Two closed paths are dependent if the multiplication of one by a set of plaquettes yields the other. This relation divides the noncontractible loops into equivalence classes, where any pair of loops in the same class are dependent. The number of these classes is $2g$. Furthermore, as discussed above, each loop can be split into two distinct logical operators. Hence, the total number of logical qubits becomes:
\begin{equation}\label{eq:logicals}
	k = 2 g = \frac{n}{8} + 2,
\end{equation}
and the resulting code's encoding rate is:
\begin{equation}
	\frac{k}{n} = \frac{1}{8} + \frac{2}{n},
\end{equation}
which is always greater than 1/8 and approaches 1/8 as ${n\rightarrow\infty}$.

Although the choice of the octagonal lattices ensures local face three-coloring, not every periodic octagonal lattice is face-three-colorable. This limitation emerges from potential inconsistencies along noncontractible closed paths. Notably, larger lattices have more independent noncontractible closed paths, complicating the search for face-three-colorable large lattices among generic octagonal lattices. However, this is not an issue for the lattices that are symmetric (arc-transitive) graphs. In such lattices, any two edges map into each other under an automorphism of the lattice. As a result, classes of equivalent (nonindependent) noncontractible loops also exhibit mutual interchangeability. Therefore, if a lattice's faces can be three-colored along a single noncontractible loop, the three-coloring can be extended to the entire lattice.

The last relevant code parameter is the code distance. During the update process in Floquet codes (Fig.~\ref{fig:logical_evolution}), each logical operator's weight (number of nonidentity Paulis) changes through six rounds, assuming its minimum value in some of the rounds. For instance, the logical operator shown in Fig.~\ref{fig:logical_evolution} has a minimum weight of four. The code distance, $d$, is the lowest minimum weight found among all logical operators. Note that with the aforementioned constructions of the logical operators, each logical operator's minimum weight, and consequently the code distance, are always even numbers for a Floquet code. 

We assess the periodic octagonal lattices from~\cite{conder2002trivalent}. Each lattice in this list that is  face-three-colorable is the basis for a hyperbolic Floquet code. As detailed above, the corresponding code parameters, ${[[n, k, d]]}$, can be calculated. Among the codes with a given distance, $d$, the one requiring the fewest physical qubits, $n$, is the most efficient, i.e., it uses the least amount of resources to achieve the same distance. Table \ref{table:code_family} lists the code parameters and lattice properties for the most efficient hyperbolic Floquet codes for even distances from 2 to 10 \footnote{The colored adjacency matrices of the hyperbolic and honeycomb lattices used for Floquet codes in this paper, as well as Python scripts for evaluating these codes are available online at \url{https://github.com/fahimniya/HyperbolicFloquetQECC}.}. The maximum code distance achieved based on octagonal lattices in~\cite{conder2002trivalent} is 10. We designate these distance-maximizing codes using the prefix `H' followed by their respective number of physical qubits.

\section{Error Models and Decoding Method} \label{sec:decoding}

\paragraph*{Error Models.}
We employ two error models. A \textit{phenomenological} model comprising unbiased depolarization errors affecting individual physical qubits and measurement errors associated with the weight-two measurements of the checks, and an \textit{entangling measurements} model which imitates circuit-level noise for future platforms with native two-qubit measurements, such as Majorana fermions~\cite{sarma_majorana_2015}. Considering other noise models and optimizing the code/decoder to the noise is an interesting avenue for future work.

The strength of the error in both models is set by a single parameter, $p$, that is a measure of the physical error rate per measurement round. In the \textit{phenomenological} error model, an unbiased depolarization channel of strength $p$ applies an X, Y, or Z Pauli operator to each physical qubit between consecutive rounds, each with an equal probability of $p/3$. Simultaneously, a measurement error channel of strength $p$ causes an erroneous outcome for each measurement with a probability $p$. In the \textit{entangling measurements} error model \cite{gidney_faulttolerant_2021}, each check measurement is accompanied by an error channel with probability $p$. In the case of an error, a non-identity weight-two Pauli is applied to the qubits of the measured check, and in half of the cases the measurement outcome is erroneous too.

Error detection properties of the hyperbolic Floquet codes are very similar to the honeycomb Floquet codes. For readability, this Section contains a summary of the essentials of our approach, and the full technical description is given in Appendix~\ref{sec:detection_and_correction}. %

Inferred values of the plaquette operators, called \textit{plaquette syndromes} or simply \textit{syndromes}, construct a ${(2+1)D}$ space-time lattice that we refer to as the \textit{syndrome lattice}. In the absence of errors, consecutive inferences of each plaquette syndrome yield the same value. Conversely, an updated plaquette syndrome that differs from its previous value implies the presence of an error (or errors) affecting that syndrome. We call these events \textit{detection events}, and they are used in identifying the occurred errors.

Each single-qubit Pauli error, depending on the type of Pauli operator and the most recent round of measurements, yields either two or four detection events. We call Pauli error with two detection events \textit{simple Pauli errors} and those with four detection events \textit{compound Pauli errors}. Two-qubit Pauli errors give rise to the detection events corresponding to their constituent single-qubit Paulis, and measurement errors are associated with four detection events. (see Fig.~\ref{fig:pauli_errors} and Fig.~\ref{fig:measurement_error}.)

To simplify our decoder, we leverage the fact that all errors can be expressed as equivalent combinations of simple Pauli errors resulting in the same detection events and impacting the logical operators identically. Specifically, our decoder operates under the assumption that only simple Pauli errors have occurred. It then identifies a set of such errors associated with the observed detection events. Given that each simple error is linked to two detection events, forming an error edge in the syndrome lattice, the decoder performs this by pairing (with a path of possible error edges) each detection event in the syndrome lattice with another detection event. To achieve this, we use the Minimum-Weight Perfect Matching (MWPM) decoder from the PyMatching package~\cite{higgott_sparse_2025}.
 
Misidentifying more complex errors as combinations of simple ones can have a negative impact on the performance of the decoder. However, being able to utilize MWPM speeds up the decoding process significantly. In this paper, we do not optimize the decoder substantially and instead focus on establishing the existence of a threshold. Since our decoder is easily adaptable to honeycomb Floquet codes, we can also perform a direct comparison between hyperbolic and honeycomb Floquet codes.

\section{Fault-Tolerant Memory Threshold} \label{sec:results}

\paragraph*{Logical Error Rates.}
With the codes, error model, and the decoder defined in Secs.~\ref{sec:hyperbolic_Floquet_code} and \ref{sec:decoding}, we can proceed to analyze how hyperbolic Floquet codes perform in storing logical qubits while being affected by the introduced error channels impacting the physical qubits.

For each hyperbolic Floquet code listed in Table~\ref{table:code_family}, this analysis involves sampling instances of decoding scenarios. In each decoding scenario, depolarization and measurements error channels with strength $p$ (as defined in Sec.~\ref{sec:decoding}) are applied during $3d$ rounds of check measurements, corresponding to $d$ Floquet periods, where $d$ is the code distance. Repetition of measurements is needed to preserve the logical memory  for a long time in the presence of measurement and decoherence processes. The threshold for further rounds of syndrome extraction (up to times polynomial in $d$) should be identical~\footnote{Appendix~\ref{sec:hd_threshold} shows some results where we substitute these $3d$ rounds with $d$ to $4d$ rounds to show that this choice does not affect the results reported here.}.  At lower depth, the threshold may shift to higher values as it becomes closer to the code-capacity threshold.

We denote individual Pauli or measurement errors as $E_i$'s. As explained in Sec.~\ref{sec:decoding} and Appendix \ref{sec:detection_and_correction}, compound single-qubit Pauli, two-qubit Pauli, and measurement errors can be decomposed as multiple simple Pauli errors. Thus, each $E_i$ is equivalent to a subset of all possible error edges in the syndrome lattice, based on its simple Pauli decomposition. This subset of error edges can be represented by a vector in the vector space of all possible error edges in the entire syndrome lattice, with each error edge corresponding to a simple Pauli error [Fig.~\ref{fig:pauli_errors}(b) and (c)]. This vector space operates over a ${\mathbb{Z}_2=\{0, 1\}}$ field, where addition is carried out modulo 2. The $\mathbb{Z}_2$ vector of each subset of error edges contains a 1 for the edges within that subset and a 0 for every other edge.
Then, the total error in the syndrome lattice, $E$, comprised of individual errors is equal to:
\begin{equation}\label{eq:errors}
	E = \sum_i E_i,
\end{equation}
where the summation is well-defined as a modulo-2 addition of the $\mathbb{Z}_2$ vectors representing individual errors ($E_i$'s).

The vertices at the two ends of an error edge in the ${(2+1)D}$ syndrome lattice represent the detection events associated with that simple Pauli error. If that error occurs, the plaquette syndromes of these vertices change sign during the round corresponding to the vertex. Consequently, a vertex that is shared by an odd (even) number of error edges within a subset of all error edges will (will not) exhibit a detection event should this subset of errors occur. The subset of syndrome lattice vertices that are shared by an odd number of error edges within a set of error edges is termed the \textit{endpoints} of that set. The endpoints of the error edges corresponding to total error, $E$, indicate the detection events linked to these errors. We denote these detection events as $D$.

The decoder's role is to find a set of errors, denoted as $C$, that would generate the same detection events, $D$, in the syndrome lattice. This set of errors is the proposed \textit{correction} by the decoder. Our MWPM decoder carries this out by finding a set of error edges with the minimum number of edges that has $D$ as their endpoints. The correction, $C$, is then the set of simple Pauli errors corresponding  to these error edges. The combined effect of the occurred errors and the decoder's correction is linked to the error edges corresponding to $E+C$. Since both the error edges of $E$ and $C$ share $D$ as their endpoints, the error edges corresponding to $E+C$ have no endpoints and form closed paths in the syndrome lattice. Given the periodicity of the syndrome lattice in its hyperbolic $2D$ plane, these ${(2+1)D}$ loops can be either contractible or noncontractible. For each noncontractible loop within the hyperbolic manifold, there exists at least one logical operator that anticommutes with the set of errors of that loop. If $E+C$ comprises an odd number of loops that anticommute with a specific logical operator, the decoding process has resulted in a logical error impacting that logical operator.

\begin{figure}[bt]
	\centering
	\includegraphics[width=0.996\linewidth]{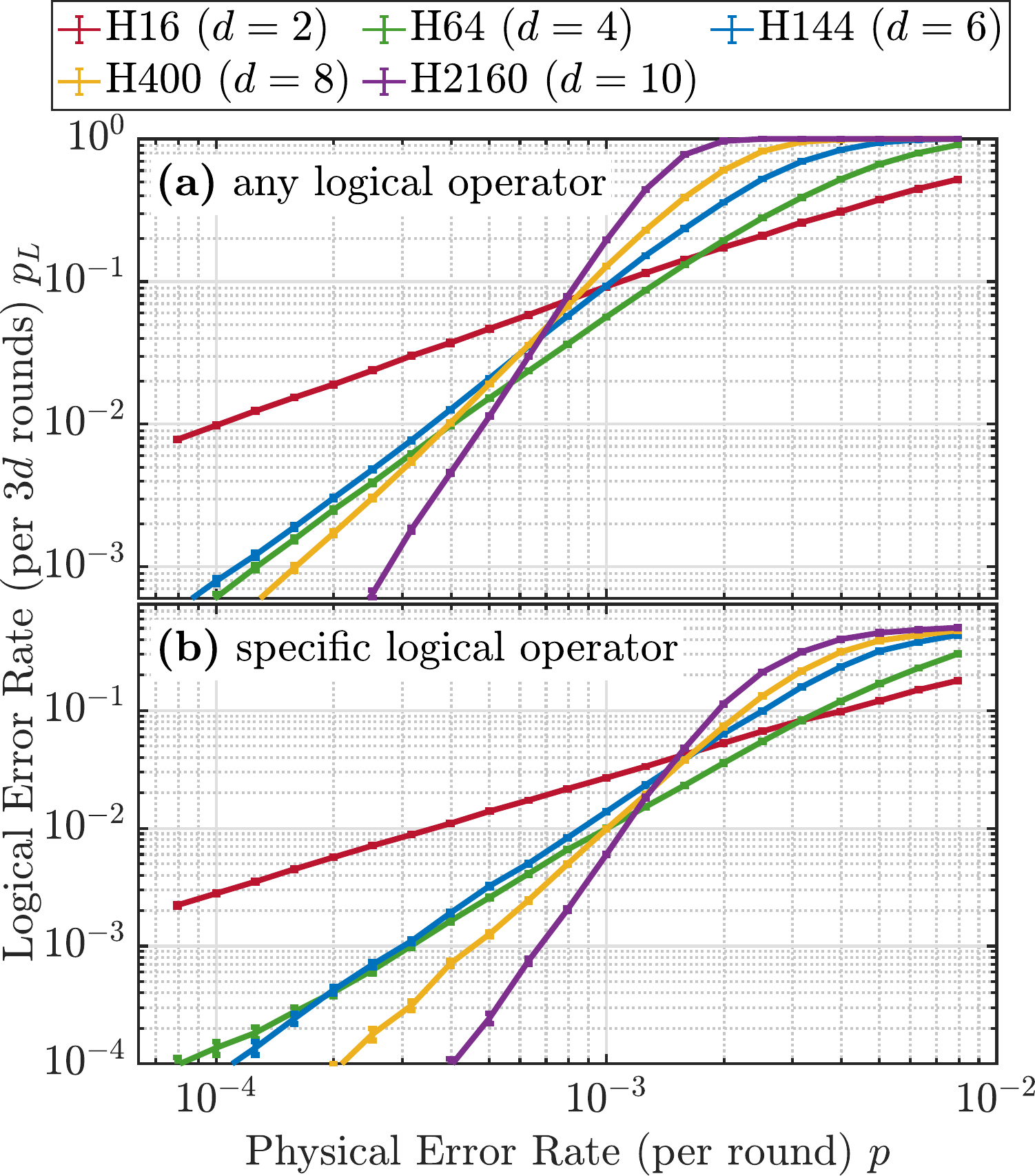}
	\caption{Logical error rates for the family of hyperbolic Floquet codes listed in Table~\ref{table:code_family} under the phenomenological error model. (a) The \textit{any logical operator} error rate. An error in any of the $2k$ logical operators of each code is considered a logical error for this rate. (b) The \textit{specific logical operator} error rate. For this rate, failure or success of the decoder is defined based on having an error or not impacting a specific logical operator within $2k$ logical operators of each code. Since hyperbolic Floquet codes are based on symmetric graphs, the choice of the specific logical operator does not matter for obtaining this plot.}
	\label{fig:logical_error_rate}
\end{figure}

\begin{figure}[tb]
    \centering
    \includegraphics[width=\linewidth]{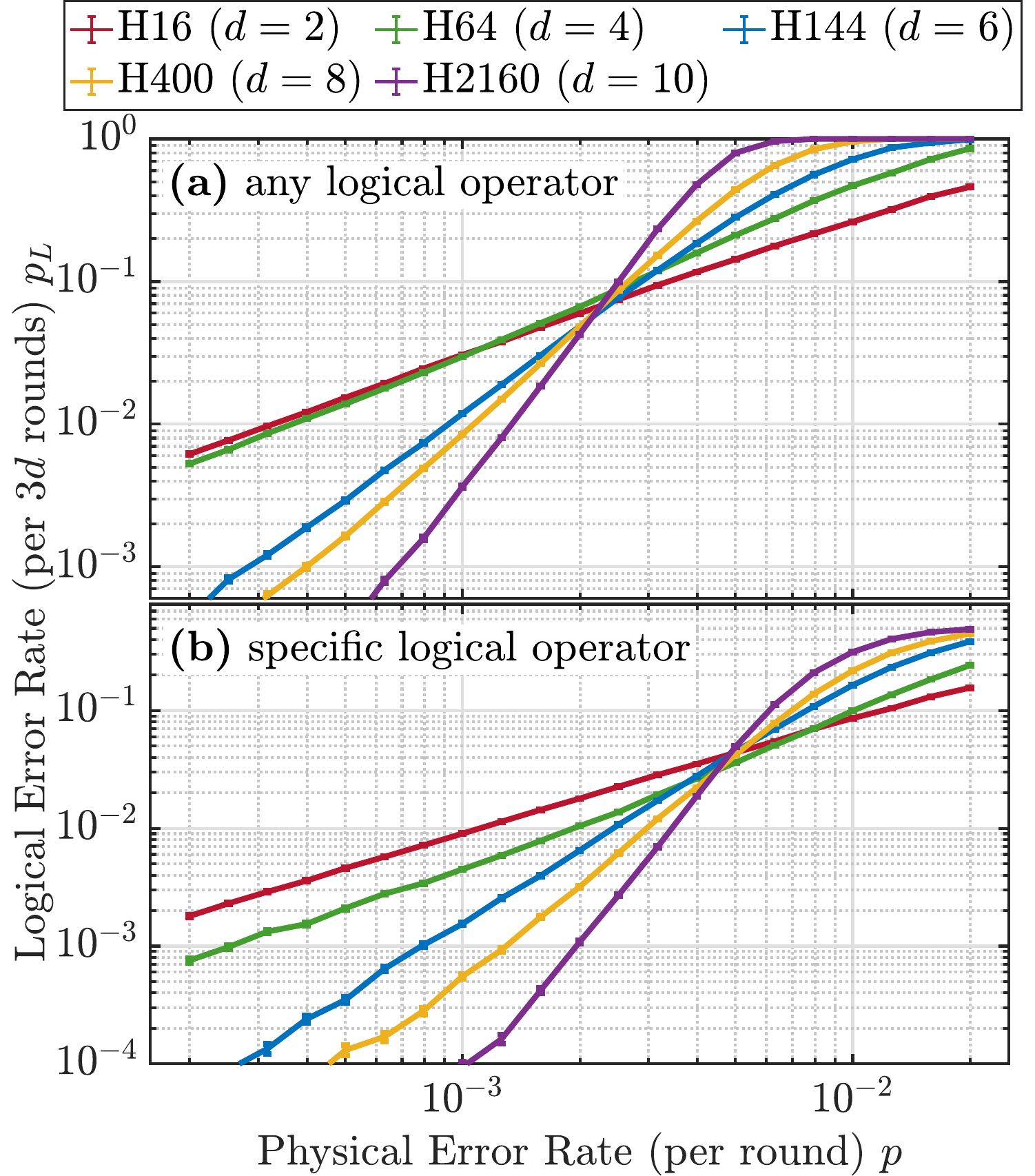}
    \caption{Logical error rates for the family of hyperbolic Floquet codes listed in Table~\ref{table:code_family} under the entangling measurements error model. (a) The \textit{any logical operator} error rate. (b) The \textit{specific logical operator} error rate. (See the caption of Fig.~\ref{fig:logical_error_rate} for additional information about the two logical error rates.)}
    \label{fig:EM_logical_error_rate}
\end{figure}

\begin{figure*}[tb]
	\centering
	\includegraphics[width=\linewidth]{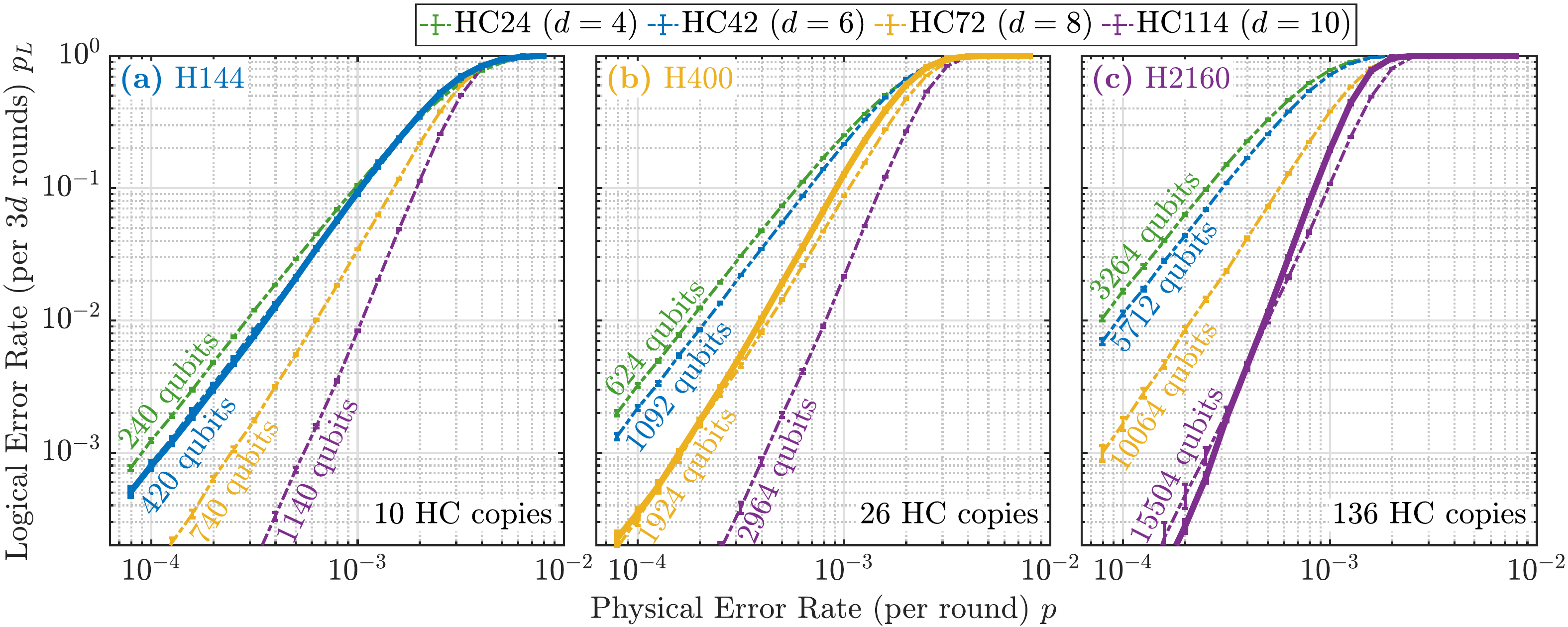}
	\caption{Comparison between hyperbolic and honeycomb Floquet codes under the phenomenological error model. Panels (a), (b), and (c) show the \textit{any logical operator} error rate of H144, H400, and H2160 hyperbolic codes (solid lines) compared to 10 copies, 26 copies, and 136 copies of honeycomb codes (dashed lines), respectively. Honeycomb codes are denoted by HC followed by their number of physical qubits. Each honeycomb code stores two physical qubits. The number of copies of each honeycomb code is chosen such that each hyperbolic code with $k$ logical qubits (see Table~\ref{table:code_family}) is compared to ensembles of honeycomb codes that also have $k$ total logical qubits. %
    The total number of physical qubits needed for that many copies of honeycomb codes is noted alongside the respective dashed line.}
	\label{fig:comparison}
\end{figure*}

Figures~\ref{fig:logical_error_rate} and \ref{fig:EM_logical_error_rate} present the MWPM decoder's performance in correcting the physical errors for the hyperbolic Floquet codes listed in Table~\ref{table:code_family} \footnotemark[2], under the \textit{phenomenological} and \textit{entangling measurements} error models, respectively. The numerically calculated logical error rates have been determined by running the decoder on the syndrome lattice after applying the physical errors over $3d$ rounds of check measurements, with $d$ the code distance\footnotemark[3]. Then we can examine whether each logical operator has been impacted by an error by counting the number of loops in $E+C$ that anticommute with it. In Fig.~\ref{fig:logical_error_rate}(a) and Fig.~\ref{fig:EM_logical_error_rate}(a), an error impacting any of the $2k$ logical operators is considered a failure in a decoding scenario. Conversely, in Fig.~\ref{fig:logical_error_rate}(b) and Fig.~\ref{fig:EM_logical_error_rate}(b), we focus on solely a specific logical operator and assess decoder's performance concerning it. Since these codes are based on symmetric graphs (as discussed in Sec.~\ref{sec:hyperbolic_Floquet_code}), the resulting logical error rate is independent of the specific equivalence class of the chosen logical operator. Hence, Fig.~\ref{fig:logical_error_rate}(b) and Fig.~\ref{fig:EM_logical_error_rate}(b) would be the same, up to numerical variations, if any other specific logical operator were chosen.

At high physical error rates, both defined logical error rates saturate at a maximum value. This saturation value is 0.5 for the \textit{specific logical operator error rate}. The reason behind this is that the combined effect of too many errors on a logical operator is equal to an unbiased coin toss: in half of scenarios, there may be an even number of noncontractible loops in $E+C$ that cross this logical operator once, hence canceling each other's effects and leaving the logical operator unaffected. For the \textit{any logical operator error rate}, however, the probability that $E+C$ has no effects on any of the $2k$ logical operators of a code is $1/2^{2k}$, which vanished as $k$ increases. Therefore, this logical error rate saturates at $\sim1$ for larger codes.

At lower values of the physical error rate, both \textit{any logical operator error rate} and \textit{specific logical operator error rate} [panels (a) and (b), respectively, in both figures] show a threshold behavior, as expected from a family of codes with increasing distance. Specifically, for sufficiently low physical error rates, ${p < p_0}$, the logical error rate decreases as the code's physical qubit count increases past $d = 4$. The threshold physical error rate, $p_0$, for both defined logical error rates is approximately $0.1\, \%$ per measurement round for the phenomenological error model (Fig.~\ref{fig:logical_error_rate}) and $0.2-0.5\, \%$ per measurement round for the entangling measurements error model (Fig.~\ref{fig:EM_logical_error_rate}). However, it is noticeably smaller for the \textit{any logical operator error rate} [Fig.~\ref{fig:logical_error_rate}(a) and Fig.~\ref{fig:EM_logical_error_rate}(a)] compared to the \textit{specific logical operator error rate} [Fig.~\ref{fig:logical_error_rate}(b) and Fig.~\ref{fig:EM_logical_error_rate}(b)]. We denote these two values as $p_0^{\rm all}$ and $p_0^{\rm one}$, respectively, and ${p_0^{\rm all} < p_0^{\rm one}}$. This subtle difference is a characteristic of finite encoding-rate codes. What links these two defined error rates is the number of logical operators, whereby a code with larger distance and a larger number of physical qubits has more logical operators that could experience a \textit{specific logical operator} error, resulting in a \textit{any logical operator} error.

The regime ${p_0^{\rm all}< p < p_0^{\rm one}}$ is an intermediate phase. For these error rates, the error rate for each logical operator is exponentially suppressed with the code distance~\cite{bravyi_simulation_2013}. On the other hand, there is an extensive number of logical operators that could potentially fail. Since the code distance is proportional to the logarithm of the number of physical qubits, and thus, of the number of logical operators, the possibilities for failure also grow exponentially with the code distance. As a result, even in the thermodynamic limit, the \textit{any logical operator error rate} can remain nonzero for these intermediate values of the physical error rate, $p$. Only when ${p < p_0^{\rm all} < p_0^{\rm one}}$, will the \textit{any logical operator error rate} also experience an exponential suppression with the code distance. This situation closely resembles the difference between the paramagnetic-ferromagnetic phase transitions in a random-bond Ising model (RBIM) on a hyperbolic lattice and in a dual-RBIM on the same lattice, as investigated in Ref.~\cite{placke_randombond_2023}.

In addition to resulting in different values for the threshold physical error rate, error models also affect the behavior of each code. For example, H64 with a code distance of 4 (defined based on the minimum weight of logical operators during Floquet rounds), behaves similarly to H144 ($d=6$) under the phenomenological error model (Fig.~\ref{fig:logical_error_rate}), while it only slightly outperforms H16 ($d=2$) under the entangling measurements error model (Fig.~\ref{fig:EM_logical_error_rate}). This behavior is not unique to the hyperbolic Floquet codes and is observed for the honeycomb Floquet codes as well (see Appendix~\ref{sec:HC_app}). This shows that the true performance of a code under an error model depends on a code distance that is defined for that specific error model~\footnote{See Appendix~\ref{sec:empirical_distance} for more details in this regard.}.

\vspace{0.1in}
\paragraph*{Comparison to Honeycomb Floquet Codes.}
The threshold physical error rates for hyperbolic codes are typically lower than those for related zero encoding-rate codes~\cite{breuckmann_constructions_2016, breuckmann_hyperbolic_2017}. This trend holds true for hyperbolic Floquet codes when compared to honeycomb Floquet codes as well. However, relying solely on the threshold value is not a useful method for comparing different codes. This value pertains to the thermodynamic limit and obscures the actual values of logical error rates, which are more practically relevant. Furthermore, below their threshold rates, hyperbolic codes are storing an extensive number of logical qubits, while the non-hyperbolic codes only preserve a constant number of them. Therefore, the threshold value alone falls short when comparing these codes performing the same task, namely, storing a given number of logical qubits. In response to this, we introduce a new figure of merit that aims at making this comparison~\footnote{Independently introduced in Refs.~\cite{bravyi_highthreshold_2024,higgott_constructions_2024} as well.}.

\begin{figure*}[tb]
    \centering
    \includegraphics[width=\linewidth]{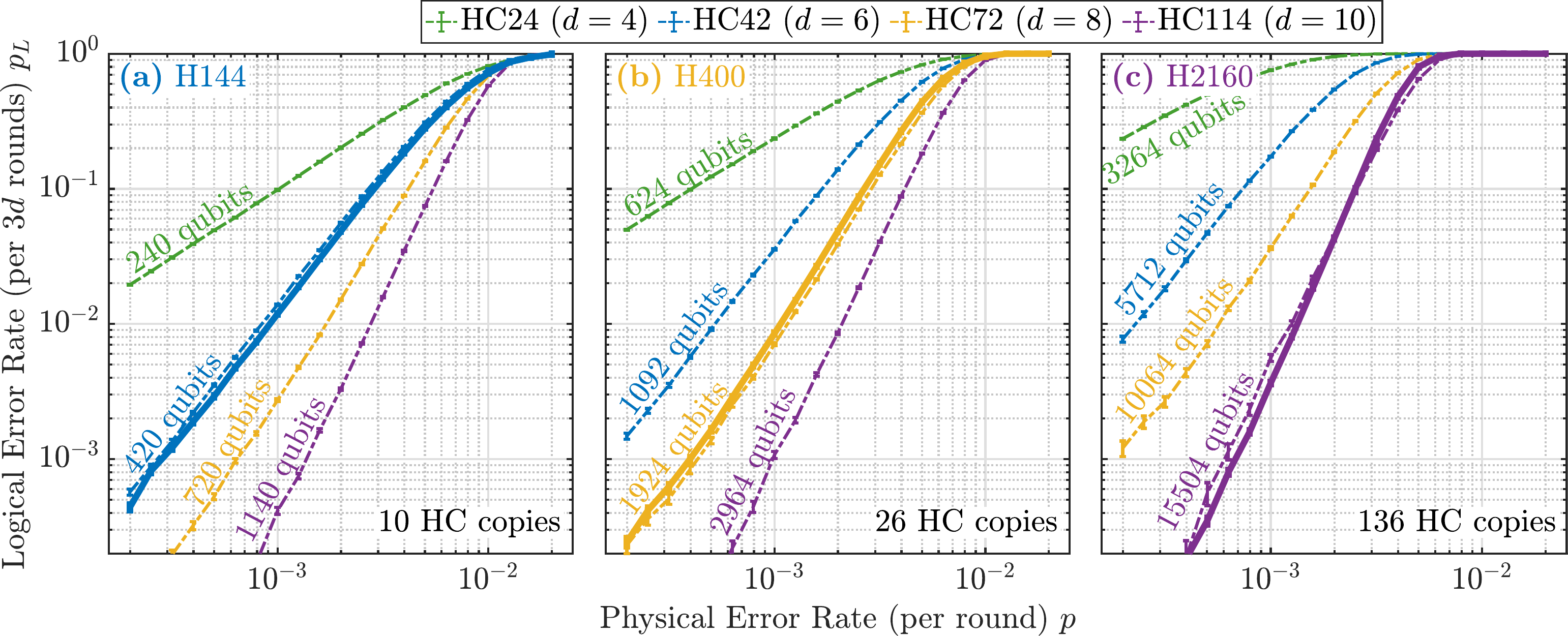}
    \caption{Comparison between hyperbolic and honeycomb Floquet codes under the entangling measurements error model. Panels (a), (b), and (c) show the \textit{any logical operator} error rate of H144, H400, and H2160 hyperbolic codes (solid lines) compared to 10 copies, 26 copies, and 136 copies of honeycomb codes (dashed lines), respectively. Honeycomb codes are denoted by HC followed by their number of physical qubits. Each honeycomb code stores two physical qubits. The number of copies of each honeycomb code is chosen such that each hyperbolic code with $k$ logical qubits (see Table~\ref{table:code_family}) is compared to ensembles of honeycomb codes that also have $k$ total logical qubits. The total number of physical qubits needed for that many copies of honeycomb codes is noted alongside the respective dashed line.}
    \label{fig:EM_comparison}
\end{figure*}

In Fig.~\ref{fig:comparison} and Fig.~\ref{fig:EM_comparison}, we present our proposed comparison between hyperbolic Floquet codes and honeycomb Floquet codes, under the phenomenological and entangling measurements error models, respectively. The four honeycomb Floquet codes featured in these plots, denoted by HC followed by their respective number of physical qubits, are based on symmetric trivalent graphs in Ref.~\cite{conder2002trivalent} with a \textit{girth} of six, and are closed hexagonal lattices tessellating the surface of a torus~\footnote{Read more about the honeycomb Floquet codes and their performance in Appendix~\ref{sec:HC_app}.}. Each panel in Fig.~\ref{fig:comparison} and Fig.~\ref{fig:EM_comparison} illustrates the \textit{any logical operator error rate} of a hyperbolic Floquet code (solid lines) compared to a number of copies of the honeycomb code with the same number of logical qubits (dashed lines). For instance, Fig.~\ref{fig:comparison}(a) and Fig.~\ref{fig:EM_comparison}(a) compare the H144 code (${[[144, 20, 6]]}$) with 10 copies of various honeycomb codes. Each of these honeycomb codes stores two logical qubits, and hence, 10 copies have a storage capacity for 20 logical qubits, equal to H144. The total number of physical qubits used in 10 copies of each honeycomb code is noted alongside its respective dashed line. The plotted logical error rate, $p_L$, for honeycomb codes is obtained by the reasonable assumption that the logical events in different copies are independent of each other.

As evident in Fig.~\ref{fig:comparison} and Fig.~\ref{fig:EM_comparison}, hyperbolic Floquet codes have performances on par with honeycomb codes with the same distance while using a significantly smaller number of physical qubits. %
For example, H400, with just 400 physical qubits, competes with honeycomb codes with up to 1924 qubits [Fig.~\ref{fig:comparison}(b) and Fig.~\ref{fig:EM_comparison}(b)], and H2160, using 2160 physical qubits, compares to 15504-qubit honeycomb codes [Fig.~\ref{fig:comparison}(c) and Fig.~\ref{fig:EM_comparison}(c)], providing respectively 5x and 7x reductions in the number of physical qubits.
In general, as the logical error rate suppression is exponential in the code distance~\cite{bravyi_simulation_2013}, it is expected that it would take honeycomb codes with equal or greater distances than a hyperbolic code to achieve a superior performance at low physical error rates. Thus, from a practical perspective, if their implementation is feasible within the architecture, hyperbolic Floquet codes represent a more sensible choice than honeycomb codes.

\section{Discussion}\label{sec:discussion}

In this work, we introduced a new family of dynamically generated codes called hyperbolic Floquet codes. These codes are useful for fault-tolerant implementation of multi-qubit quantum memories utilizing weight-two measurements. The hyperbolic nature of these codes enables us to find codes with a finite encoding rate and an increasing distance with the total number of physical qubits. We investigated the fault-tolerance of these codes under two different error models by testing an MWPM decoder on them and showing the existence of a threshold error rate, below which the logical error rates are suppressed. By comparing to honeycomb Floquet codes, we made the case that, given the same number of physical qubits, hyperbolic Floquet codes can achieve logical error suppression better than copies of honeycomb codes storing the same number of logical qubits. However, a noteworthy disadvantage for the hyperbolic codes is that the decoding task for a hyperbolic code-block of a given number of logical qubits can be more time-consuming than decoding copies of honeycomb code-blocks encoding the same number of logical qubits. Improving and optimizing decoders for hyperbolic codes, similar to other quantum LDPC codes, could reduce this barrier.

When encoding $k$ logical qubits into $n$ data qubits, we may also use $c$ auxiliary qubits to measure the error syndrome. This means the code uses a total of $n+c$ physical qubits. Thus, the net encoding rate can be defined as ${r = k/(n+c)}$~\cite{bravyi_highthreshold_2024}.
For traditional surface code architectures, $c = n-1$,
whereas for the newer quantum LDPC codes, $c=n$~\cite{bravyi_highthreshold_2024}.
In our scheme, which takes inspiration from the properties of the Floquet code, we have $c=0$ for platforms with native two-body measurements, leading to resource savings. Consequently, our net encoding rate is identical to the regular encoding rate, that is ${k/n = 1/8 + 2/n}$.

To further improve the logical error suppression of our approach, one can interpolate between hyperbolic and honeycomb Floquet codes by moving beyond regular hyperbolic tessellations. This has been previously implemented for hyperbolic surface codes, resulting in the so-called \textit{semi-hyperbolic surface codes}~\cite{breuckmann_hyperbolic_2017}. The infusion of additional qubits within each regular polygon of a hyperbolic Floquet code will result in a code with the same number of logical qubits, but higher number of physical qubits. This new code, however, will have a greater distance than its parent. This procedure, sometimes referred to as \textit{leap-frogging}, can be carried out repeatedly. Therefore, there exist families of codes with lower encoding rates than our hyperbolic Floquet codes, but a more favorable distance, although still scaling logarithmically with the number of physical qubits. Each family of these codes can also have higher threshold error rates than regular hyperbolic Floquet codes, approaching the threshold error rate of the honeycomb codes.

The imperfections in its components challenge the building of a quantum computer. For example, some qubits might possess manufacturing defects, rendering them ``dead.'' These dead qubits are either entirely nonfunctional or have a noise level substantially higher than their counterparts. For the surface code, there are several existing strategies~\cite{stace_thresholds_2009,stace_error_2010,tang_robust_2016,nagayama_surface_2017,strikis_quantum_2023,siegel_adaptive_2023} to address these dead qubits. A recent method, as mentioned in Ref.~\cite{aasen_faulttolerant_2023}, suggests recoupling the lattice applicable to any code that operates on a face three-colorable lattice. The recoupling can be viewed as a re-triangulation of the three-colorable lattice, ensuring dead qubits aren't utilized in the subsequent measurement schedule. Since our codes are also based on face three-colorable lattices, we can use the aforementioned strategy from Ref.~\cite{aasen_faulttolerant_2023} for handling dead qubits. 

In this work, we have focused on the fault-tolerant memory threshold of hyperbolic Floquet codes. For an error correcting code family to be useful for quantum computation it is also necessary to implement state preparation, logical gates, and measurement. State preparation and measurement protocols for our hyperbolic Floquet codes can follow standard protocols from honeycomb Floquet codes~\cite{hastings_dynamically_2021,gidney_faulttolerant_2021,magdalenadelafuente_$mathrmxyz$_2025}. Efficiently implementing logical gates with limited time overhead for finite-rate LDPC codes remains a fundamental challenge~\cite{gottesman_faulttolerant_2014}.  However, there has been recent progress adapting lattice surgery methods to LDPC codes that could be applied to our setting~\cite{breuckmann_hyperbolic_2017,horsman_surface_2012,Cohen_2022,xu_constantoverhead_2024a}.

\paragraph*{Note added:} 
While this manuscript was in preparation, Ref.~\cite{higgott_constructions_2024} appeared discussing hyperbolic Floquet codes using a slightly different lattice construction method. A threshold was reported for zero-rate hyperbolic codes using the \textit{leap-frogging} method. 
Owing to the difference in lattice constructions, they fail to capture some of the distance-maximizing codes presented here.

\begin{acknowledgements}
    We thank I.\ Boettcher, H.\ Nguyen, Y.-Z.~Chou, A.\ Lavasani, V.\ Albert, and M.\ Barkeshli for useful discussions. This research was supported in part by NSF QLCI (award No.~OMA-2120757), the Air Force Office of Scientific Research, and the Simons Foundation. A.F., K.B., and A.V.G.~were also supported in part by the DoE ASCR Accelerated Research in Quantum Computing program (award No.~DE-SC0020312),  DoE ASCR Quantum Testbed Pathfinder program (award No.~DE-SC0019040), NSF PFCQC program (award No.~1818914), AFOSR (awards No.~FA95502010108 and No.~FA9550-19-1-0275), ARO MURI (award No.~W911NF1610349), AFOSR MURI (award No.~FA9550-20-1-0323), and DARPA SAVaNT ADVENT (award No.~W911NF2120106). H.D. acknowledges support from DOE DE-AC02-05CH11231, AFOSR FA9550-22-1-0339, OMA-2120757 and the Simons Investigator in Physics Award. 
    Support is also acknowledged from the U.S.~Department of Energy, Office of Science, National Quantum Information Science Research Centers, Quantum Systems Accelerator.  The authors acknowledge the University of Maryland supercomputing resources (\url{http://hpcc.umd.edu}) made available for conducting the research reported in this paper.
\end{acknowledgements}

\bibliographystyle{quantum}
\bibliography{references.bib}

\clearpage
\onecolumngrid

\appendix

\section{Details of Error Detection and Correction}\label{sec:detection_and_correction}

Sec.~\ref{sec:decoding} contains a summary of the essentials of our decoding method. In this Appendix, we provide the full technical description of error detection and correction.

\begin{figure}[h]
	\centering
	\includegraphics[width=0.9\linewidth]{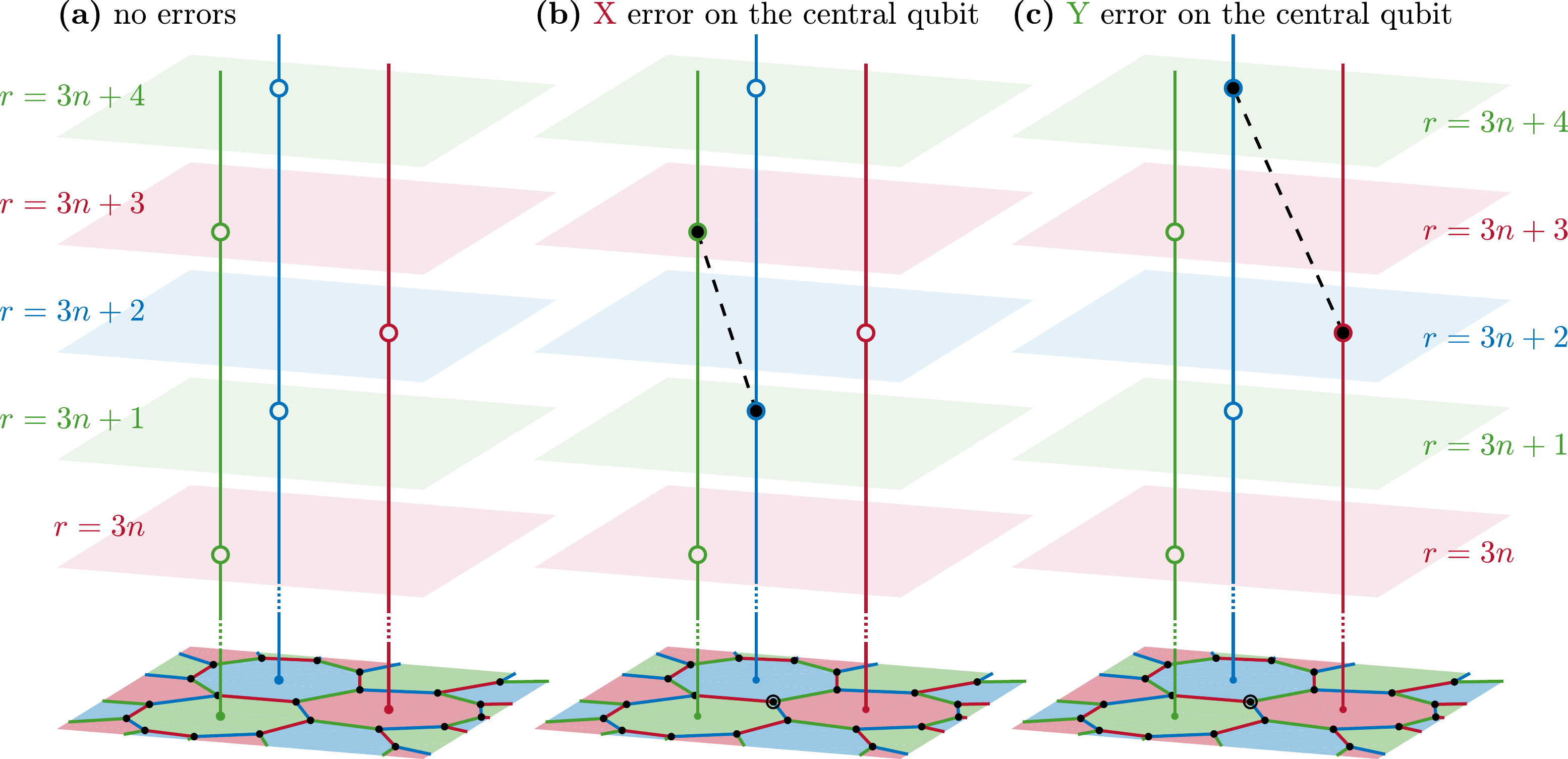}
	\caption{Detection events for Pauli errors. (a) ${(2+1)D}$ lattice of plaquette syndromes. Each plaquette syndrome is inferred every third measurement round, denoted by empty circles in the timelines of the plaquettes. (b) X and (c) Y Pauli errors between rounds ${r=3n}$ and ${r=3n+1}$ lead to detection events marked on lines perpendicular to the hyperbolic plane. In the presence of errors, inference of a plaquette syndrome may yield a value different than its most recent inferred value. In this case, the inferred syndrome exhibits a detection event, denoted by a filled circle. For the Pauli errors happening between rounds ${r=3n}$ and ${r=3n+1}$, X and Y Paulis are simple Pauli errors and are associated with two detection events, connected to each other by an error edge in the syndrome lattice. A Z Pauli, on the other hand, is a compound Pauli error and is equivalent to a combination of these X and Y Paulis encompassing all four detection events. The four detection events of a compound Pauli error form a hyper-edge in the syndrome lattice.}
	\label{fig:pauli_errors}
\end{figure}

\paragraph*{Syndrome Lattice.}
Floquet codes possess the capability to detect potential error occurrences and provide the necessary information for attempting to correct the effects of these errors on logical qubits. Error detection is executed by utilizing the inferred values of the plaquette operators, hereafter referred to as \textit{plaquette syndromes} or simply \textit{syndromes}. Each plaquette syndrome is inferred every third round. The inference rounds depend on plaquette's color; specifically, the \colorX, \colorY, and \colorZ{} plaquette syndromes are inferred at rounds $r=3n+2$, $3n+3$, and $3n+1$, respectively, for every nonnegative integer $n$. (No plaquette syndrome is inferred at round ${r=0}$.) By extending a line perpendicular to the hyperbolic plane for each plaquette and marking each syndrome update with a circle on this line, we construct a ${(2+1)D}$ lattice comprising updates of plaquette syndromes, as partially depicted in Fig.~\ref{fig:pauli_errors}(a). We refer to this lattice as the \textit{syndrome lattice}.

In the absence of errors, the plaquette operators become part of the ISG upon their initial inference (rounds $r=2$, $r=3$, and $r=1$ for \colorX, \colorY, and \colorZ{} plaquettes, respectively) and remain in the ISG, as they commute with all check operators. Consequently, any subsequent inference of a plaquette syndrome yields a value equal to the previously inferred one. This condition is denoted by empty circles in the ${(2+1)D}$ lattice of syndromes in Fig.~\ref{fig:pauli_errors}(a). Conversely, an updated plaquette syndrome that differs from its previous value implies the presence of an error (or errors) affecting that syndrome. Following Gidney, we call these events \textit{detection events}. They are indicated by filled circles in the ${(2+1)D}$ lattice of syndromes.

\vspace{0.1in}
\paragraph*{Error Syndromes.}
The lattice's edge three-coloring ensures that the three measurement checks involving a given physical qubit have distinct Pauli operators associated with that qubit. As a result, a Pauli error on a physical qubit anticommutes with two of those checks. For instance, the X Pauli error on the qubit marked with a circle in Fig.~\ref{fig:pauli_errors}(b) anticommutes with the \colorY{} (YY) and \colorZ{} (ZZ) checks intersecting at that vertex. Assuming the error occurs between rounds ${r=3n}$ and ${r=3n+1}$, the immediately following \colorY{} and \colorZ{} checks will be affected by the error. The \colorY{} check's faulty outcome will be used in the inference of the adjacent \colorZ{} plaquette syndrome resulting in a detection event at round ${r=3n+1}$. This is denoted by a filled circle in Fig.~\ref{fig:pauli_errors}(b). At round ${r=3n+2}$, the adjacent \colorX{} plaquette syndrome will be inferred using two faulty outcomes of both \colorY{} and \colorZ{} checks, and hence, will be agnostic to the error. No detection event for the \colorX{} plaquette syndrome is marked with an empty circle in Fig.~\ref{fig:pauli_errors}(b). The \colorY{} plaquette syndrome's inference at round ${r=3n+3}$, however, will use the outcomes from the faulty \colorZ{} check and the unaffected \colorX{} check, exhibiting a detection event similar to the \colorZ{} plaquette syndrome, indicated with a filled circle. The two detection events of the X Pauli error %
form an error edge in the syndrome lattice, represented by a dashed line in Fig.~\ref{fig:pauli_errors}(b).

The analysis for a Y Pauli error on the same qubit follows similarly. The Y Pauli error on the qubit marked with a circle and its detection events are illustrated in Fig.~\ref{fig:pauli_errors}(c). The Y Pauli anticommutes with the \colorX{} and \colorZ{} checks that intersect at the vertex, while leaving the \colorY{} check unaffected. Assuming once again that the error occurs between rounds ${r=3n}$ and ${r=3n+1}$, the \colorY{} check's measurement at round ${r=3n+1}$ is agnostic to the error. However, the \colorZ{} check's and the \colorX{} check's measurements at rounds ${r=3n+2}$ and ${r=3n+3}$ do capture the error. During round ${r=3n+2}$, the \colorX{} syndrome is updated using the outcomes of the unaffected \colorY{} check and the faulty \colorZ{} check, thereby triggering a detection event. In the next round ${r=3n+3}$, the \colorY{} syndrome is updated using faulty outcomes from both the \colorZ{} and \colorX{} checks, leaving it unaffected. Lastly, at round ${r=3n+4}$, the outcomes from the faulty \colorX{} check and the unaffected \colorY{} check lead the \colorZ{} syndrome to exhibit another detection event. These two detection events of the Y Pauli error %
form an edge in the syndrome lattice, as shown in Fig.~\ref{fig:pauli_errors}(c). Notably, for these errors occurring between rounds ${r=3n}$ and ${r=3n+1}$, the two detection events for the Y Pauli error happen with a one round delay as compared to the case of the X Pauli error. 

A Z Pauli error that occurs between rounds ${r=3n}$ and ${r=3n+1}$ can be regarded as a combination of X and Y Pauli errors, following from ${Z=-iXY}$. As a result, the associated detection events are also a combination of those corresponding to X and Y Pauli errors. Specifically, a Z Pauli error between rounds ${r=3n}$ and ${r=3n+1}$ leads to the following four detection events: the \colorZ{} syndrome at ${r=3n+1}$, the \colorX{} syndrome at ${r=3n+2}$, the \colorY{} syndrome at ${r=3n+3}$, and once again, the \colorZ{} syndrome at ${r=3n+4}$. These four detection events create a hyper-edge (i.e., an ``edge'' that connects more than two vertices) in the syndrome lattice.

The decisive factor for a Pauli error to yield either two or four detection events is the most recent round of measurements. In the outlined example above, the Pauli errors take place between rounds ${r=3n}$ (XX measurements) and ${r=3n+1}$ (YY measurements). In this scenario, X and Y Pauli errors generate two detection events, while the Z Pauli is associated with four. We call Pauli errors with two detection events (X and Y in this case) \textit{simple Pauli errors} and those with four detection events (Z in this case) \textit{compound Pauli errors}. During the interval between YY and ZZ measurements, Y and Z Pauli errors are simple, while X Pauli errors are compound. Similarly, between ZZ and XX measurements, Z and X Pauli errors are simple, and Y Pauli errors are compound. 

\begin{figure}[bt]
	\centering
	\includegraphics[width=0.55\linewidth]{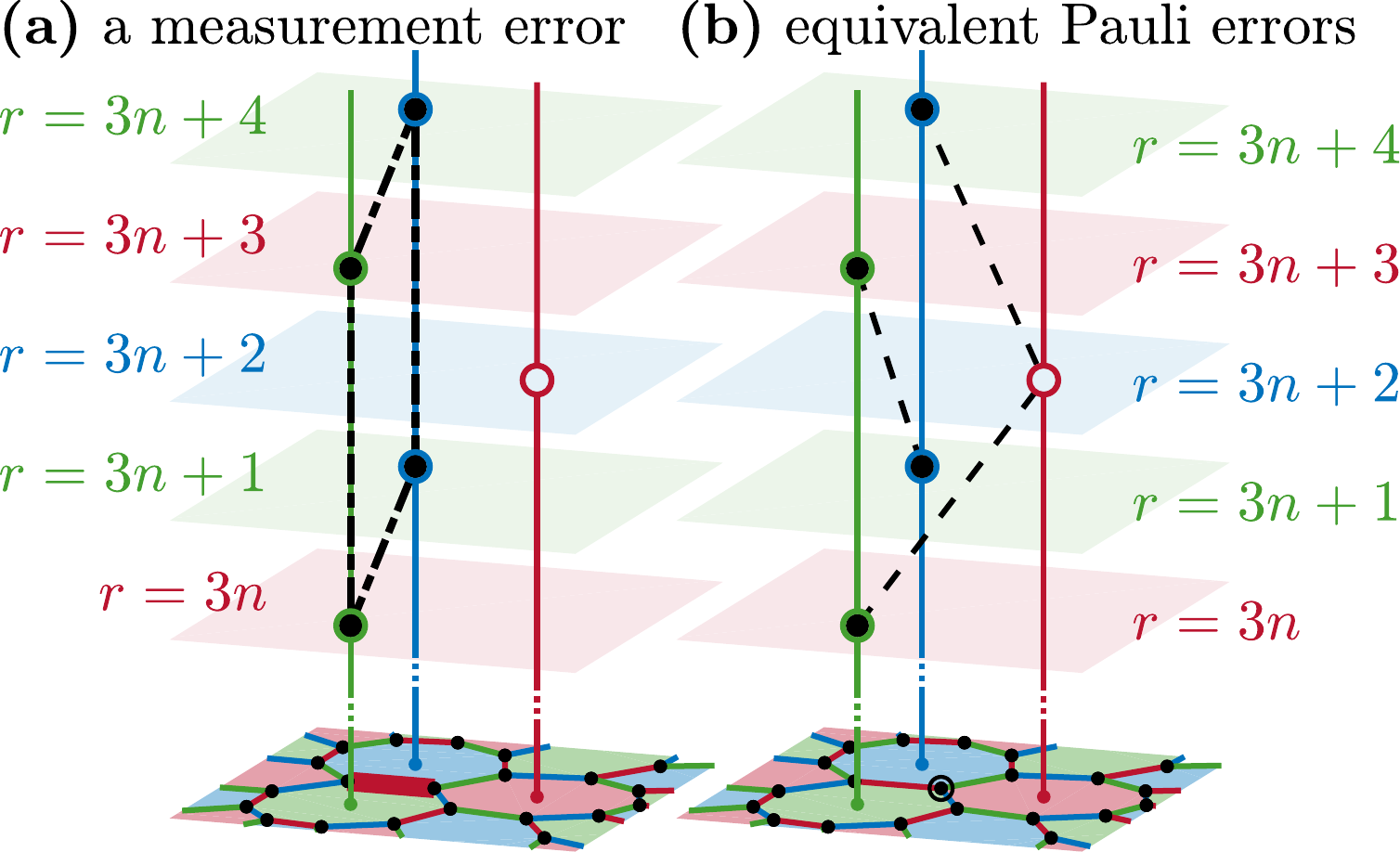}
	\caption{Detection events of a measurement error. (a) The faulty measurement outcome of the highlighted \colorX{} check at round ${r=3n}$ is used to infer an incorrect value for the adjacent \colorY{} and \colorZ{} plaquette syndromes at rounds ${r=3n}$ and ${r=3n+1}$. Later correct measurement of the same check leads to correct inference of these syndromes three rounds later. In both cases, the inferred syndrome differs from the most recent inference and gives rise to a detection event. Hence, a measurement error has four detection events forming a hyper-edge in the syndrome lattice, shown with a dashed parallelogram. (b) A combination of three simple Pauli errors on one of the qubits of the faulty check, denoted with a circle, gives rise to the same detection events. Three simple Pauli errors in this case are: a Z error between rounds ${r=3n-1}$ and ${r=3n}$, an X error between rounds ${r=3n}$ and ${r=3n+1}$, and a Y error between rounds ${r=3n+1}$ and ${r=3n+2}$. These equivalent Pauli errors have the same effects on the logical operators as the measurement error.}
	\label{fig:measurement_error}
\end{figure}

In a measurement error, the outcome of a check's measurement is faulty. This faulty outcome is then utilized in inferring two syndromes, causing them to exhibit detection events. An example is illustrated in Fig.~\ref{fig:measurement_error}(a). In this example, the highlighted \colorX{} check's outcome at round ${r=3n}$ is erroneous, while all other measurements are accurate. This faulty \colorX{} check leads to the \colorY{} plaquette syndrome at round ${r=3n}$ and the \colorZ{} plaquette syndrome at round ${r=3n+1}$ exhibiting detection events. Given the assumption that all other measurements are accurate, the next measurement of the same \colorX{} check at round ${r=3n+3}$ relates to the accurate value of the plaquette syndromes. Thus, the same \colorY{} and \colorZ{} syndromes generate additional detection events three rounds later. A measurement error, therefore, is linked to four detection events, forming a hyper-edge in the syndrome lattice, depicted with a dashed parallelogram in Fig.~\ref{fig:measurement_error}(a).

\vspace{0.1in}
\paragraph*{Error Detection and Correction.}
The occurrence of each error may affect the logical operators of the code. For a Pauli error on a physical qubit, this happens when that qubit is on the noncontractible closed path of a logical operator featuring a distinct nonidentity Pauli on it compared to the error. In this scenario, the Pauli operator and the logical operator anticommute, leading the error to induce a sign flip in the logical operator. In the case of a measurement error, it impacts a logical operator if the measured check is along the closed path of that logical operator. Consequently, the faulty measurement outcome is used in updating the logical operator's expression, resulting in an incorrect sign. We argue that the detection events associated with each error provide sufficient information for attempting to correct the respective error. This attempt, however, is not always successful, and if the number of errors equals to or exceeds $d/2$, can lead to logical errors.

Consider the pair of detection events linked to a simple Pauli error. There are four different Pauli errors that could trigger these two detection events and, thus, are indistinguishable from one another. Without any loss of generality, assume the simple Pauli error is an X error. The two plaquette syndromes with detection events correspond to adjacent \colorY{} and \colorZ{} octagons [see Fig.~\ref{fig:pauli_errors}(b)]. These two octagons identify the location of the physical qubit affected by the X error to their shared \colorX{} edge, without distinguishing which of the two ends of that edge it is. We label these two qubits as $q_1$ and $q_2$. Moreover, the X error could have occurred either between the most recent XX and YY measurements or between the most recent ZZ and XX measurements. These possibilities for error's support qubit, $q_1$ or $q_2$, and its timing, between ZZ and XX measurements or between XX and YY measurements, give rise to the four indistinguishable errors with identical detection events.

Now consider a logical operator, $L$, that contains one of these two qubits ($q_1$, without loss of generality) in its closed path. Between the most recent ZZ and XX measurements, $L$'s expression can either be type one (Z Paulis on \colorX{} checks) or type two (X Paulis on \colorZ{} checks). In each case, between the most recent XX and YY measurements, $L$'s expression will be type two (Y Paulis on \colorX{} checks) or type one (X Paulis on \colorY{} checks), respectively. The X error on $q_1$ affects $L$ only if $L$'s expression features Y or Z Paulis on $q_1$ at the time of the error. This holds true if $L$ is a type one and then a type two logical operator, in addition to having the \colorX{} edge of $q_1$'s vertex along its closed path. In this case, $L$'s expression during the two possible time intervals includes Y or Z Paulis on both $q_1$ and $q_2$. Therefore, all four indistinguishable simple Pauli errors impact logical operators in the same way, and their indistinguishability does not pose an obstruction in correcting their effects on logical operators.
 
A compound Pauli error is indistinguishable from the pair of simple Pauli errors that constitute its detection events. In the  example illustrated in Fig.~\ref{fig:pauli_errors}, a Z error on the qubit marked with a circle between rounds ${r=3n}$ and ${r=3n+1}$ causes the same detection events as a combination of X and Y errors on that qubit occurring during the same time interval (or their indistinguishable counterparts). Let us label this qubit $q_1$ and consider a logical operator, $L$, with $q_1$ on its path. Between XX and YY measurements, $L$'s expression is either type one (X Paulis on \colorY{} checks) or type two (Y Paulis on \colorX{} checks). The Z error affects $L$ only if its expression features X or Y Paulis on $q_1$. In such a case, one of the X or Y errors on $q_1$ anticommutes with $L$, while the other one commutes with it. As as result, a compound Pauli error and its indistinguishable simple Pauli error pairs impact logical operators in the same way, and the indistinguishability again does not pose an obstruction to error correction.
 
The indistinguishability analysis also extends to measurement errors. A measurement error is indistinguishable from a combination of three simple Pauli errors. These equivalent Pauli errors for an erroneous measurement of a \colorX{} check are shown in Fig.~\ref{fig:measurement_error}(b). The three Pauli errors consist of: first, a Z between ${r=3n-1}$ and ${r=3n}$; second, an X between ${r=3n}$ and ${r=3n+1}$; and third, a Y between ${r=3n+1}$ and ${r=3n+2}$. All these errors impact the physical qubit marked with a circle in Fig.~\ref{fig:measurement_error}(b). Equivalent Pauli errors for measurement errors of central \colorY{} and \colorZ{} checks can be obtained by $2\pi/3$ or $4\pi/3$ rotations in the hyperbolic plane, followed by one- or two-round shifts in time.
 
A logical operator, $L$, with the faulty \colorX{} edge on its closed path will incorporate the erroneous outcome of its measurement in the update protocol. This will lead to an incorrect sign in its expression. If $L$'s expression immediately before the measurement error at round $r=3n$ is type one (Z Paulis on \colorX{} checks), its subsequent two expressions will be type two (Y Paulis on \colorX{} checks) and, then, type one (Y Paulis on \colorZ{} checks). Among the three equivalent Pauli errors, the first and third one commute with $L$, while the second one anticommutes with it. Hence, their combined effect would be a sign flip in $L$'s expression, exactly as in the case of  the measurement error.
 
Analogously, if $L$'s initial expression immediately before the faulty measurement's round is type two (X Paulis on \colorZ{} checks), its subsequent two expressions will be X Paulis on \colorY{} checks and Z Paulis on \colorY{} checks. Depending on whether $L$'s path traverses the \colorZ{} or \colorY{} check connected to the marked qubit, either the first or third Pauli error among the three equivalent errors will anticommute with $L$, while the other two will commute with it. Therefore, the combined effect of these equivalent Pauli errors is again the same as for the measurement error. Thus, the indistinguishability of measurement errors and combinations of simple Pauli errors does not pose an obstruction to error correction.
 
To simplify our decoder, we leverage the fact that all errors can be expressed as equivalent combinations of simple Pauli errors resulting in the same detection events and impacting the logical operators identically. Specifically, our decoder operates under the assumption that only simple Pauli errors have occurred. It then identifies a set of such errors associated with the observed detection events. Given that each simple error is linked to two detection events, forming an error edge in the syndrome lattice, the decoder performs this by pairing (with a path of possible error edges) each detection event in the syndrome lattice with another detection event. To achieve this, we use the Minimum-Weight Perfect Matching (MWPM) decoder from the PyMatching package~\cite{higgott_sparse_2025}.
 
Misidentifying more complex errors as combinations of simple ones can have a negative impact on the performance of the decoder. However, being able to utilize MWPM speeds up the decoding process significantly. In this paper, we do not optimize the decoder substantially and instead focus on establishing the existence of a threshold. Since our decoder is easily adaptable to honeycomb Floquet codes, we can also perform a direct comparison between hyperbolic and honeycomb Floquet codes. 

\section{Threshold Plots with Various Duration of the Applied Errors}\label{sec:hd_threshold}

In this Appendix, we numerically study the dependence of the threshold behaviour reported in Sec.~\ref{sec:results} on the choice of $3d$ rounds of erroneous simulation for each code, where $d$ is the code distance.

\begin{figure}[h]
    \centering
    \includegraphics[width=6in]{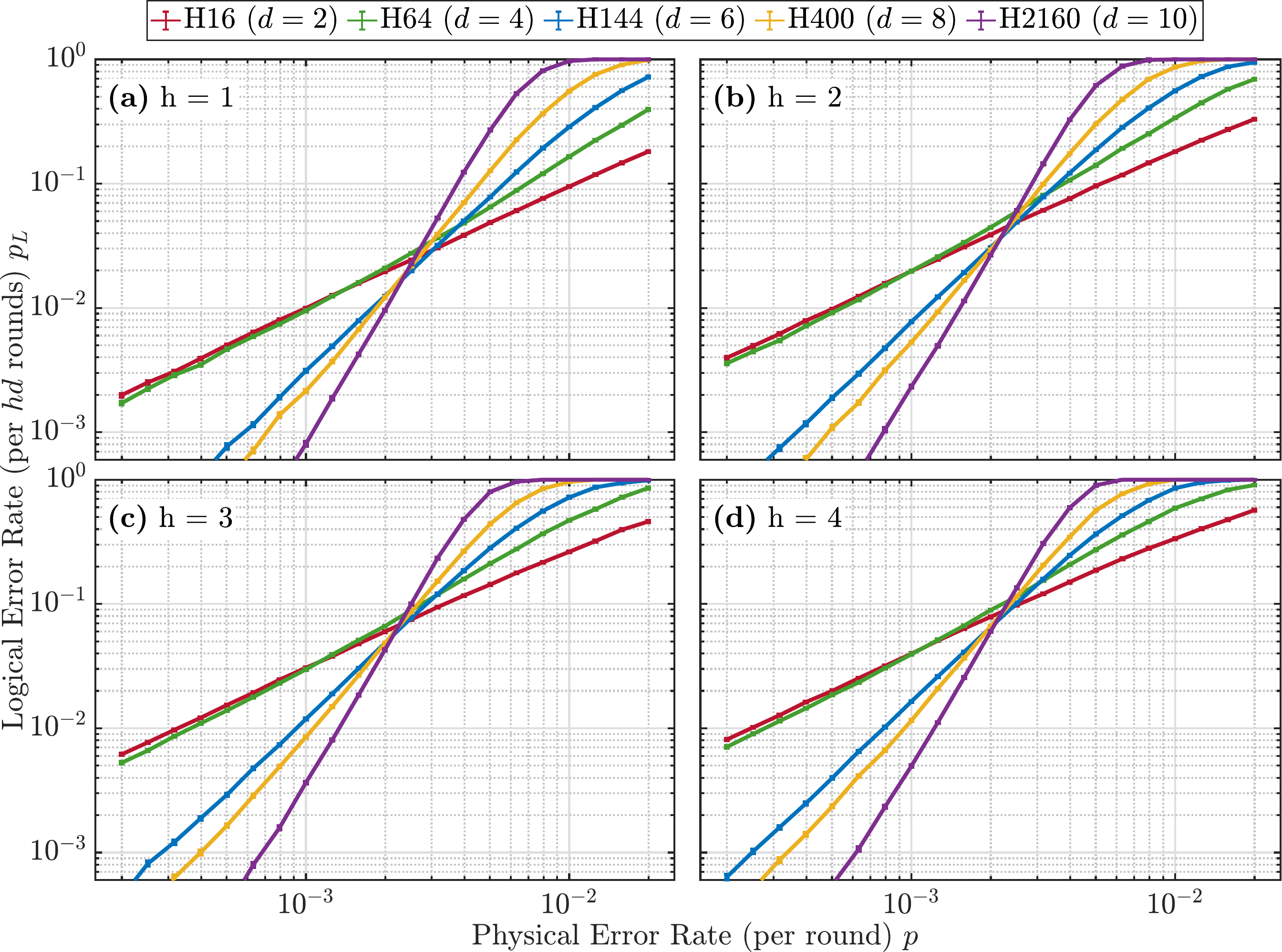}
    \caption{Logical error rates of the hyperbolic Floquet codes under the entangling measurements error model for various duration of the applied errors.}
    \label{fig:hd_threshold}
\end{figure}

In Fig.~\ref{fig:hd_threshold} we present logical error rates of the hyperbolic Floquet codes under the entangling measurement error model for various duration of the applied errors. In all four panels the duration in which the errors are applied is proportional to the distance of the code so that the observed phase transition is a property of the three-dimensional space-time lattice of syndromes. To be more precise, this duration equals $hd$ rounds of check measurements with $h=1$-$4$ for panels (a)-(d). As seen here, the approximate value of the threshold error rate is around $0.25\, \%$ for all four panels and does not depend on the value of $h$. All plots in the main text are produced with $h=3$.

\section{Empirical Distances of Hyperbolic Floquet Codes}\label{sec:empirical_distance}

In this appendix, we investigate the performance of hyperbolic codes in suppressing logical errors. It is expected that at low physical error rates, $p$, the logical error rate of a code, $p_L$, is expected to decay exponentially with the code distance, $d$,~\cite{bravyi_simulation_2013}:
\begin{equation}\label{eq:exponential_decay}
    p_L = A(d) p^{-\alpha d},
\end{equation}
where $A(d)$ is a coefficient dependent on the code distance and $\alpha$ is a constant.

\begin{figure}
    \centering
    \includegraphics[width=4in]{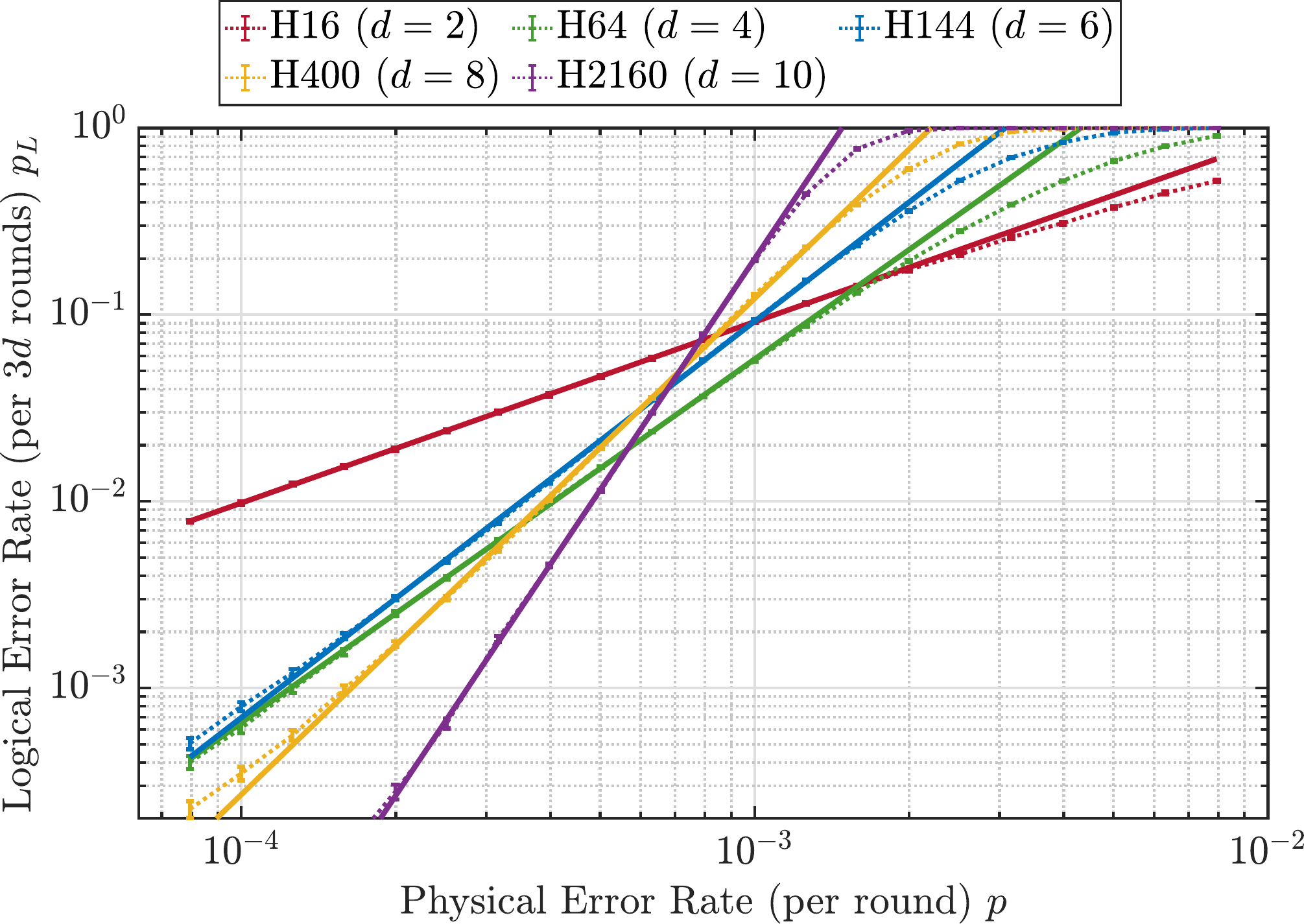}
    \caption{Any logical operator error rates for the family of hyperbolic Floquet codes under the phenomenological error model (dotted lines) [same as Fig.~\ref{fig:logical_error_rate}(a)] and linear fits to the values below the approximate error rate threshold of $p_0^{\rm all} \simeq 8 \times10^{-4}$ per round (solid lines).}
    \label{fig:logical_fit_lines}
\end{figure}

\begin{figure}[!b]
    \centering
    \includegraphics[width=4in]{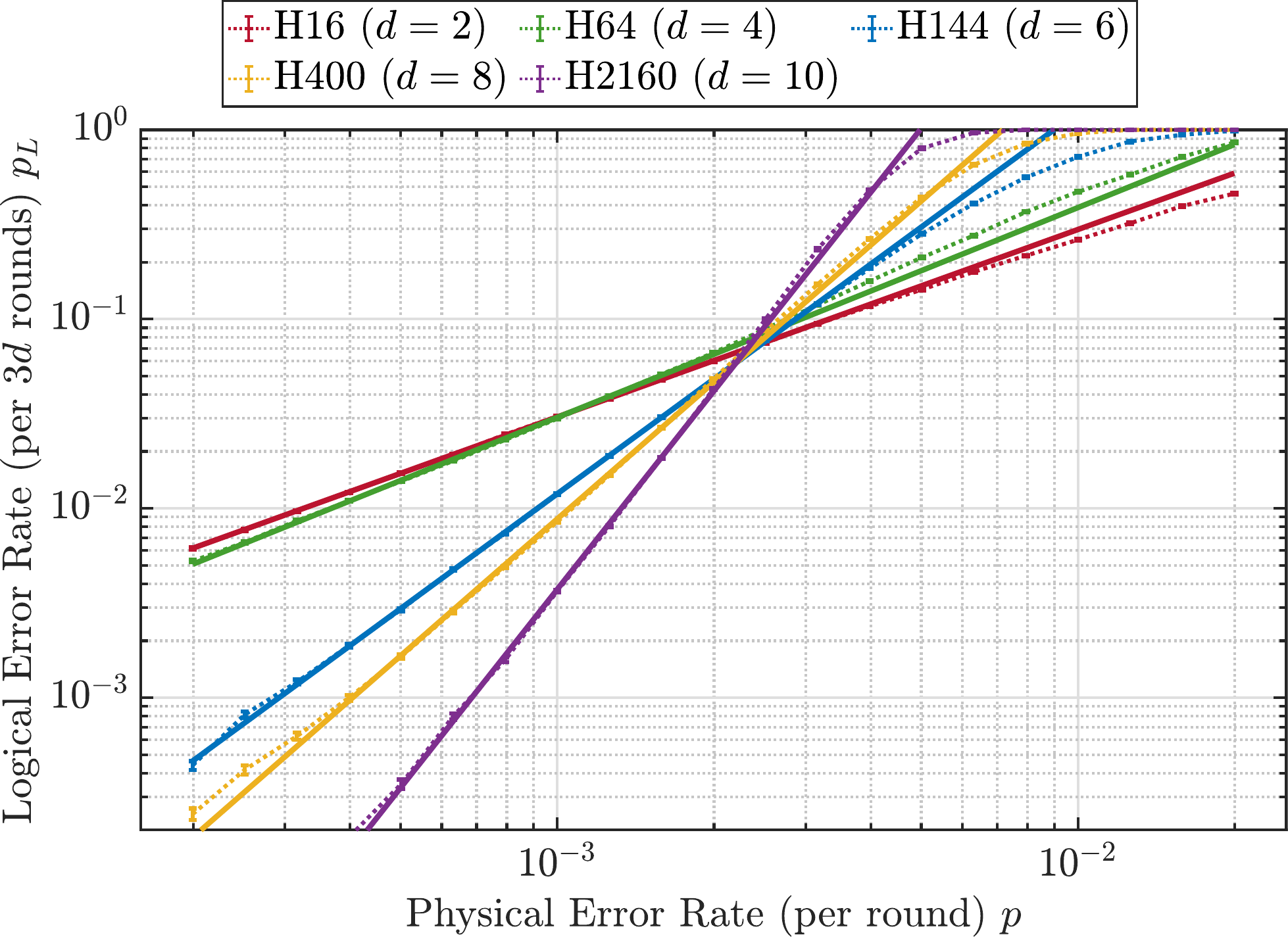}
    \caption{Any logical operator error rates for the family of hyperbolic Floquet codes under the entangling measurements error model (dotted lines) [same as Fig.~\ref{fig:EM_logical_error_rate}(a)] and linear fits to the values below the approximate error rate threshold of $p_0^{\rm all} \simeq 1.5 \times10^{-3}$ per round (solid lines).}
    \label{fig:logical_fit_lines_EM3}
\end{figure}

The linear behavior of logical error rates in log-log plots under both error models (Figs.~\ref{fig:logical_fit_lines} and \ref{fig:logical_fit_lines_EM3}) provides numerical evidence for Eq.~\ref{eq:exponential_decay}. In these figures, we also show linear fits for each line below the corresponding approximate error rate threshold. The extracted slopes from these fits should be proportional to what we call the \textit{empirical distances} of the codes. Similar to the threshold rate, we expect  empirical distances to depend on the code, the error model, and the decoder.

\begin{figure}
    \centering
    \includegraphics[width=5.5in]{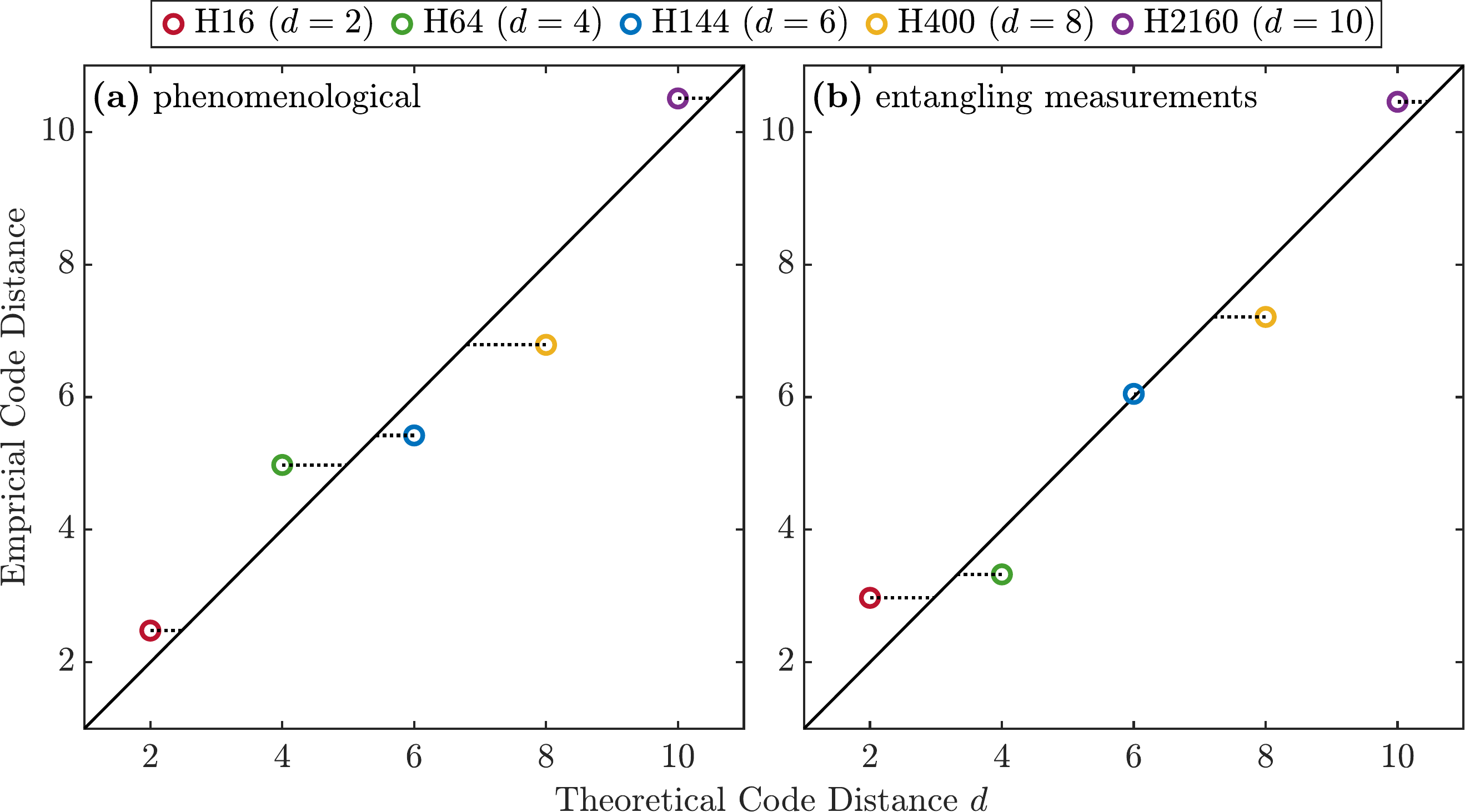}
    \caption{Empirical distances of hyperbolic codes under (a) phenomenological, and (b) entangling measurements error models. Theoretical code distances are defined as the lowest minimum weights of logical operators of a code (see Sec.~\ref{sec:hyperbolic_Floquet_code}). The empirical distance is proportional to the slope of a line fitted to the logical error rates of the codes below the approximate threshold value (see Figs.~\ref{fig:logical_fit_lines} and \ref{fig:logical_fit_lines_EM3}). If a code's data point is to the left (right) of the diagonal line with a slope equal to 1, then it is outperforming (underperforming) compared to the average of the code family. For example, H64 outperforms its theoretical code distance of $d=4$ under the phenomenological error model but underperforms under the entangling measurements error model.}
    \label{fig:empirical_distances}
\end{figure}

Figure~\ref{fig:empirical_distances} shows the obtained empirical distances of hyperbolic codes under the phenomenological (panel a) and entangling measurements (panel b) error models. These empirical distances are assumed to be proportional to the slope of the fitted lines in Figs.~\ref{fig:logical_fit_lines} and \ref{fig:logical_fit_lines_EM3}. The proportionality coefficient is chosen such that, in each of the panels in Fig.~\ref{fig:empirical_distances}, the square sum of the length of the dotted lines is minimized. These dotted lines connect each code's data point to the solid line with a slope equal to 1. The closer a data point is to this line, the better the theoretical code distance predicts that code's logical performance.

Whenever a code's data-point falls to the left of the unit-slope solid line, the code is performing better in preserving logical qubits than predicted by its theoretical code distance. We anecdotally observe that while H16 and H2160 outperform their theoretical distances under both error models, and H144 and H400 underperform under both error models, the behavior of H64 changes with the error model. Specifically, H64 outperforms its theoretical code distance of $d=4$ under the phenomenological error model, with an empirical code distance closer to that of H144 ($d=6$), but underperforms under the entangling measurements error model, behaving similarly to H16 ($d=2$). 

\section{Honeycomb Floquet Codes from Symmetric Graphs}\label{sec:HC_app}

This Appendix includes some additional information about the honeycomb Floquet codes that are used as a baseline of comparison for the performance of hyperbolic Floquet codes in Sec.~\ref{sec:results}.

These honeycomb Floquet codes are based on symmetric trivalent graphs in Ref.~\cite{conder2002trivalent} with a \textit{girth} of six, which are closed hexagonal lattices tessellating the surface of a torus. This is unlike the usual method of constructing honeycomb codes which uses rectangular patches of a hexagonal lattice with periodic boundary conditions~\cite{gidney_faulttolerant_2021}. Turns out using symmetric graphs is beneficial is terms of number of physical qubits for a given code distance. For example, codes with code distances of 8, 12, 16, and 20 based on the construction in Ref.~\cite{gidney_faulttolerant_2021} have 96, 216, 384, and 600 physical qubits, respectively. While codes with the same distance based on symmetric graphs need 72, 162, 288, and 450 physical qubits, respectively. This corresponds to a $25\, \%$ save on the resources. Using symmetric graphs for non-Floquet hexagonal codes can also lead to reducing the number of physical qubits for a given distance and make the realization of higher-distance codes in the near future more feasible.

\begin{figure}[h]
    \centering
    \includegraphics[width=4in]{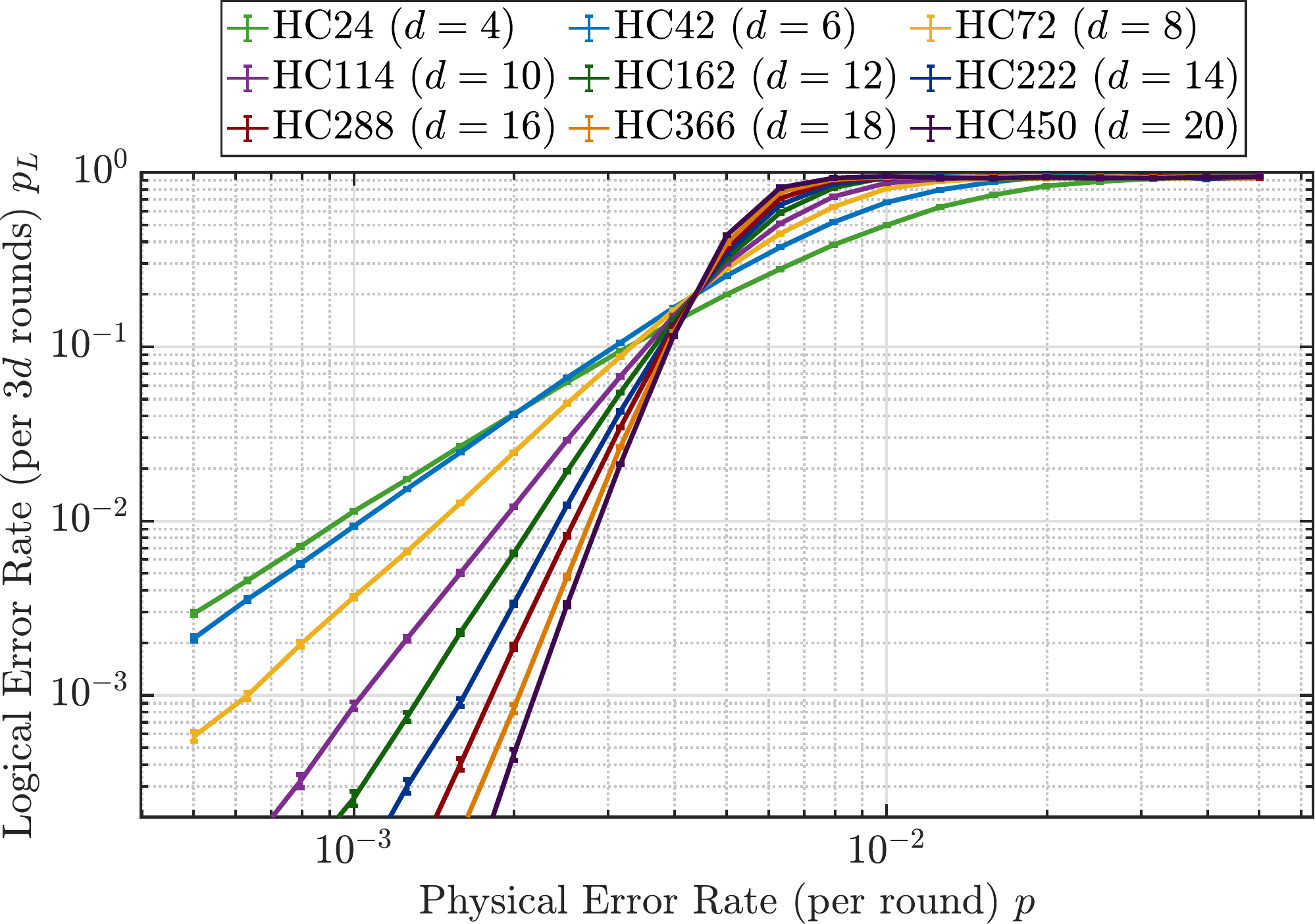}
    \caption{\textit{Any logical operator} error rates for the family of honeycomb Floquet codes with the phenomenological error model.}\label{fig:hc_threshold}
\end{figure}

\begin{figure}[h]
    \centering
    \includegraphics[width=4in]{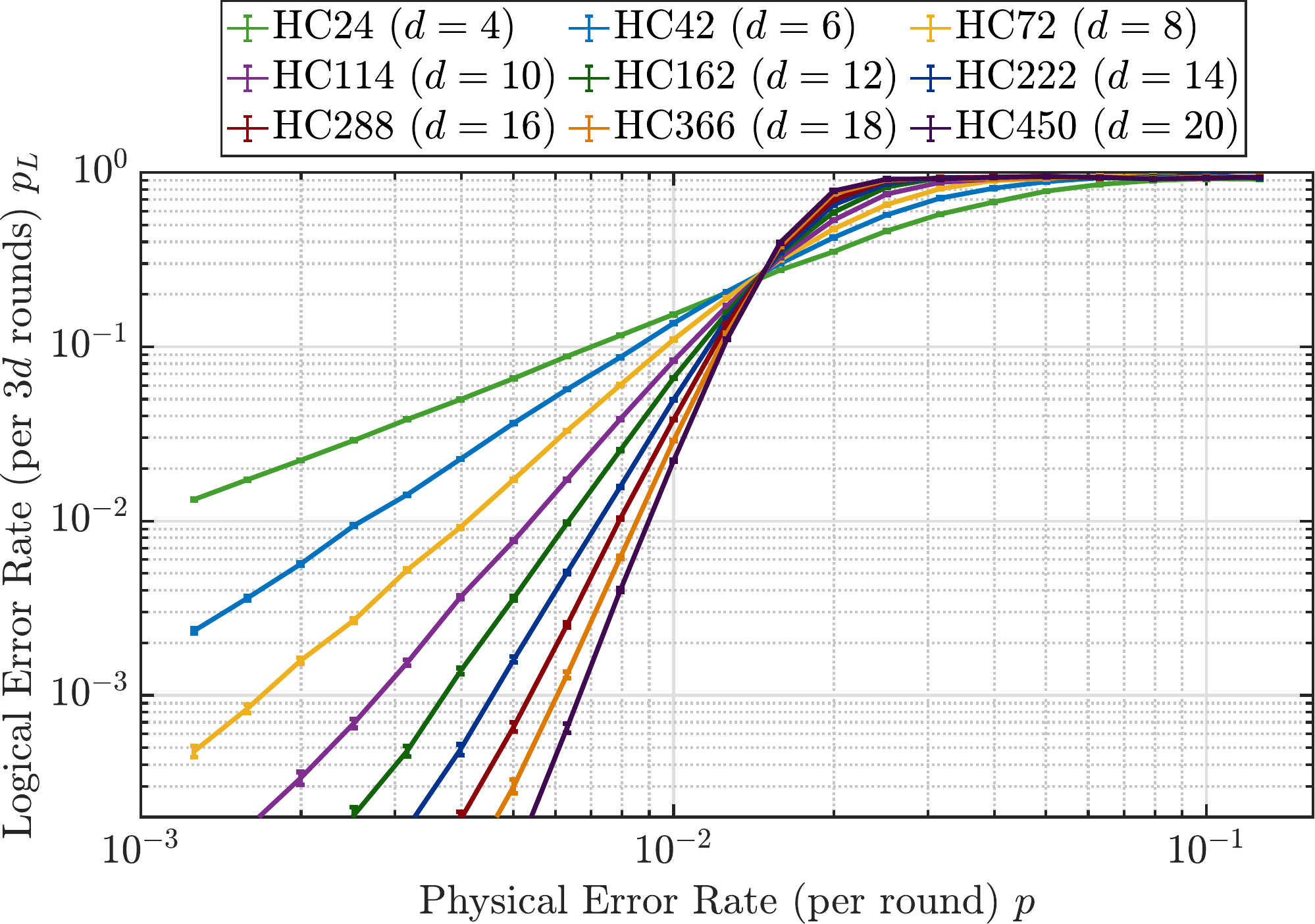}
    \caption{\textit{Any logical operator} error rates for the family of honeycomb Floquet codes with the entangling measurements error model.}\label{fig:EM_hc_threshold}
\end{figure}

In Fig.~\ref{fig:hc_threshold} and Fig.~\ref{fig:EM_hc_threshold}, we present the performance of honeycomb Floquet codes under both of our error models and decoded by the MWPM decoder described in Sec.~\ref{sec:decoding} and Appendix~\ref{sec:detection_and_correction}. Honeycomb codes show threshold error rates around $0.45\, \%$ and $1.5\, \%$ under the phenomenological and entangling measurements error models, respectively. Similar to the hyperbolic codes, the value of the threshold is higher for the entangling measurements error model.

\end{document}